\documentclass[10pt,aps,pra,preprint]{revtex4-1}
\usepackage[T1]{fontenc}
\usepackage[utf8]{inputenc}
\usepackage{amsmath}
\usepackage{amssymb}
\usepackage{bm}
\usepackage{graphicx}
\graphicspath{{images_updated_may25/}}
\usepackage{dcolumn}
\usepackage{subcaption}
\usepackage{hyperref}
\usepackage{comment}

\begin{document}

\title{Transverse Optomechanical Interaction Mediated by Mechanically Induced Symmetry Breaking: Hamiltonian Dynamics}

\author{Satyam S. Jha}
\affiliation{Department of Physics, Queens College of CUNY, Flushing, NY 11367}
\affiliation{Graduate Center of CUNY, 365 5th Ave, New York, NY 10016}

\author{Lev Deych}
\thanks{Author to whom correspondence should be addressed: lev.deych@qc.cuny.edu}
\affiliation{Department of Physics, Queens College of CUNY, Flushing, NY 11367}
\affiliation{Graduate Center of CUNY, 365 5th Ave, New York, NY 10016}

\begin{abstract}
In cavity optomechanics, the interaction between light and motion is usually introduced via the shift of cavity resonances in response to mechanical displacement. Here we present an analysis of Hamiltonian dynamics of an optomechanical system with a different form of optomechanical coupling, in which mechanical motion dynamically couples otherwise independent optical modes. In the language of Schwinger pseudospin operators, the dispersive coupling can be interpreted as ``longitudinal'' while the mode-coupling mechanism corresponds to a transverse interaction. The latter is well known in cavity and circuit QED but was given only scarce attention in cavity optomechanics. Unlike the traditional dispersive/dissipative coupling, the mode-coupling optomechanical interaction generates rich Hamiltonian dynamics even in the absence of external drive or dissipation. For instance, under certain initial conditions this dynamics is characterized by a Hamiltonian Hopf bifurcation controlled by the total photon power injected into the system. Below the bifurcation threshold and for large enough non-linearity, mechanical modulation of optical amplitudes generates a broad spectrum of multiple sidebands covering a frequency interval larger than ten mechanical frequencies. Above the threshold, the frequency of optical oscillations becomes dependent on the mechanical amplitude, while mechanical degrees of freedom return to oscillating at their bare frequency. The scope of this work is limited to the study of purely Hamiltonian dynamics to demonstrate that the mechanically mediated mode-coupling optomechanical interaction provides an alternative method of coherent control of energy exchange between light and mechanical motion.
\end{abstract}

\maketitle
\section{ Optomechanics and symmetry}\label{sec:Introduction}

Traditionally, cavity optomechanics deals with the interaction between light confined to an optical cavity and mechanical motion mediated by radiation pressure. It is often described in terms of shifts of the cavity resonant frequency and/or its decay rate in response to mechanical displacements \cite{AspelmeyerRMP}. In its conventional realization, cavity optomechanics involves optical cavity modes coupled to a mechanical oscillator and driven by an external pump establishing a steady-state intracavity photon population. Most of the optomechanical phenomena result from linear oscillations of optical and mechanical degrees of freedom around this steady state, which are described either classically or quantum-mechanically \cite{Marquardt2008,WilsonRae2007,Aspelmeyer2014}.

Cavity optomechanics based on dispersive and dissipative coupling mechanisms is by now a mature field with well developed  theoretical foundation \cite{Marquardt2008,WilsonRae2007,Aspelmeyer2014} and a number of significant experimental successes, such as  realization of strong coherent coupling between an optical cavity mode and a mechanical mode\cite{Verhagen2012Nature},  cavity-assisted cooling of mechanical oscillators to their quantum ground state\cite{Schliesser2008,Chan2011,Teufel2011,Peterson2016,Delic2020}, as well as the observation of dynamical backaction leading to self-sustained oscillations and parametric instability\cite{Braginsky2001,Marquardt2006,Ludwig2008}. The same general paradigm was also used to analyze more complex multimode and hybrid systems and diverse phenomena such as wavelength conversion and optomechanically induced dark-mode phenomena \cite{Dong2015NSR,Andrews2014NatPhys}, synchronization and collective dynamics \cite{Zhang2012PRL,Heinrich2011PRL}, optomechanically-induced transparency\cite{Weis2010Science}, and nonreciprocal photon transport \cite{Kim2015NatPhys,Xu2020Nature,Fang2017NatPhys,Miri2017}. 

In this paper we introduce an alternative optomechanical interaction mechanism that may open new venues for coherent control of the exchange of energy between mechanical and optical degrees of freedom.  This mechanism relies on mechanically induced dynamic mixing of otherwise independent optical modes, arising when mechanical motion breaks the underlying symmetry of the optical cavity. While models with mechanically induced coupling between cavity modes have been considered previously, most notably in Ref.~\cite{Cheung2011,Stannigel2012PRL,SafaviNaeini2011PRL}, the physics responsible for the coupling in our model is fundamentally different. The authors of Ref.~\cite{Stannigel2012PRL,SafaviNaeini2011PRL} dealt explicitly with optical modes belonging to two different cavities, and the configuration involving a moving membrane inside a Fabry-Perot resonator considered in Ref.~\cite{Cheung2011} also essentially divides a single cavity in two coupled ones. In essence, the mechanical oscillations in Ref.~\cite{Cheung2011,Stannigel2012PRL} modulate preexisting coupling between optical modes determined by the architecture of the corresponding optical systems. In contrast, the symmetry breaking mechanism introduced here leads to intrinsic hybridization of the optical normal modes that vanishes in the absence of the symmetry breaking mechanical displacement.  

In multimode optical resonators, the optical dynamics can naturally be expressed in terms of the Schwinger pseudospin formalism, in which dispersive or dissipative interactions act along the pseudospin longitudinal axis and may therefore be interpreted as \emph{longitudinal coupling}. By contrast, the mechanically induced intermode coupling within the same formalism, corresponds to a \textit{transverse} optomechanical interaction. Dynamics of systems with such interactions are well studied in cavity and circuit QED \cite{Blais2021RMP,Devoret2013Science,Wallraff2004Nature,Gerry2005Book,Scully1997Book}, but have not previously been considered in optomechanics. 

For instance, the authors of Ref.~\cite{Cheung2011} limited their scope to the derivation of the Hamiltonian with the transverse interaction, but discussed its dynamical consequences only in terms of a time-dependent perturbation causing transitions between initial optical modes. Authors of Ref.\cite{Stannigel2012PRL,SafaviNaeini2011PRL} focused on driven optomechanical systems in which the mechanically induced intermode coupling modifies the dynamics around externally established optical steady states. At the same time, the mechanically induced transverse coupling between cavity modes results in nontrivial Hamiltonian dynamics which, to the best of our knowledge, has not been analyzed previously.

We show in this paper that mechanically induced mode mixing gives rise to rich conservative dynamics characterized by such effects as the generation of multiple optical sidebands spanning a broad frequency range and optical frequencies that depend on \emph{mechanical} amplitudes. Additional features include Hamiltonian Hopf bifurcations\cite{Haller1991} arising for specific initial conditions and Landau–Zener–type breakdowns of adiabaticity\cite{Landau1932,Zener1932,Stueckelberg1932,Majorana1932} manifesting as abrupt transfer of optical excitation between different cavity modes. Importantly, all these effects arise in the absence of external pumping. Such a regime is not normally considered in standard optomechanical settings because canonical models with longitudinal coupling generally do not exhibit nontrivial dynamics without an external pump\cite{Clerk2020}.

These features establish the conservative dynamics of symmetry-breaking optomechanical systems as an important object of study in its own right, suggest new possibilities for applications beyond conventional dispersive optomechanical architectures, and provide a natural starting point for future studies of such systems in the presence of external pumping.

Although dissipation is unavoidable in real experiments, the dynamical effects predicted by the conservative analysis can still be accessed through short-pulse or ring-down excitations, where the evolution remains approximately Hamiltonian over experimentally accessible transient timescales comparable to or shorter than the cavity lifetime. Similar approaches are widely used in other areas of physics. For example, coherent atom–cavity dynamics in cavity QED is routinely observed within the photon lifetime of the resonator \cite{Brune1996}, while transient dynamics of weakly damped nonlinear resonators is commonly used to probe the phase-space structure and bifurcations of underlying Hamiltonian systems \cite{Almog2007}. The conservative analysis presented here therefore constitutes a necessary and self-contained first step, establishing the phase-space structure and bifurcation landscape onto which dissipative effects will be systematically introduced in subsequent work. To illustrate the robustness of the predicted phenomena, we include in the Supplementary Materials numerical simulations for a range of optical and mechanical damping rates.

\section{Model}\label{sec:model}
In its simplest form the standard optomechanical model is described by a Hamiltonian~\cite{AspelmeyerRMP}
\begin{equation}\label{eq:standard-model-H}
\hat{H}=\hbar\omega_{cav}\hat{a}^{\dagger}\hat{a}+\hbar\Omega\hat{b}^{\dagger}\hat{b}+\hbar g\hat{a}^{\dagger}\hat{a}\left(\hat{b}^{\dagger}+\hat{b}\right)  
\end{equation}
where the first term describes a single mode of an optical resonator characterized by creation-annihilation operators $\hat{a}^{\dagger},\hat{a}$, the second term corresponds to a mechanical oscillator ($\hat{b}^{\dagger},\hat{b}$ are mechanical creation annihilation operators, respectively), and the last term introduces the coupling between the mechanical and optical modes. In models with dispersive coupling, parameter $g$ is defined in terms of the derivative of the optical frequency with respect to the mechanical displacement $x$ : 
\begin{equation}\label{eq:disp_coupling}
    g=x_{zpf}\frac{\partial\omega_{cav}}{\partial x}
\end{equation}
  where $x_{zpf}$ is the zero-point fluctuation amplitude of the mechanical oscillator (see details in \cite{AspelmeyerRMP}). 

The transverse optomechanical interaction introduced in this paper may arise in multimode optical cavities whose modes are characterized by the underlying cavity symmetry. It relies on coupling between these modes mediated by mechanical motion rather than relying primarily on the dependence of the resonator frequency or decay rate on mechanical displacement.  If mechanical displacement breaks the underlying cavity symmetry, it leads to mixing of previously independent optical modes. The strength of this coupling is proportional to the mechanical displacement. 

The mechanical subsystem in our model is represented by a two-dimensional isotropic harmonic oscillator. While transverse optomechanical coupling can also arise in systems with effectively one-dimensional mechanical motion~\cite{Cheung2011}, the two-dimensional oscillator introduced here is essential for the symmetry-based structure of our model. The two orthogonal mechanical degrees of freedom are related differently to the cavity symmetry. One coordinate, denoted by $q_z$, preserves the spatial symmetry of the optical resonator and therefore gives rise primarily to conventional dispersive coupling, while the orthogonal coordinate $q_y$ breaks this symmetry and dynamically induces coupling between otherwise independent optical modes. The interplay of these two coordinates allows optical frequency shifts and intermode coupling strength to emerge as independent dynamical variables associated with distinct symmetry properties of the mechanical motion.

The optical cavity is modeled as a three-mode system: one mode with frequency $\omega_0$ and two degenerate modes with frequency $\omega_1$. This choice is motivated by a whispering-gallery-mode (WGM) spherical resonator interacting with a polarizable dipole, which we use as a paradigmatic physical realization of our model. The minimal realization of the mechanism discussed below requires only two optical modes, but the third mode provides an additional useful ``knob'' for controlling the system properties and is experimentally relevant for the WGM--dipole configuration~\cite{Deych2009, Rubin2011PRA, Rubin2011, Rubin2010}.

Such a resonator--particle system preserves axial symmetry about the line connecting the resonator center and the particle. Using this line as the polar $q_z$-axis of a spherical coordinate system, the optical modes can still be labeled by their azimuthal quantum number $m$, since coupling between modes with different orbital numbers $l$ can be neglected due to their large frequency separation. Displacements of the particle perpendicular to the $q_z$-axis then induce coupling between modes with different $m$, a direct consequence of the rotation properties of spherical harmonics. Physically, the coupling between modes described by $g_y$ reflects the fact that the nanoparticle not only shifts the resonance frequencies of the WGMs but also perturbs the spatial field distribution of the optical modes~\cite{Deych2009,Rubin2010}. When the particle moves, it drags the deformed field pattern, producing mechanically induced mode mixing that corresponds to the symmetry-breaking term of our Hamiltonian. Since the electromagnetic field of a dipole contains multipoles with azimuthal numbers $m=-1,0,1$, only three WGM resonator modes are affected by the interaction~\cite{Deych2009}.

Experimentally this setup can be implemented using a high-sphericity droplet that supports high-$Q$ WGMs. The droplet can be stabilized in a liquid or acoustic trap and coupled to light through a tapered fiber \cite{Dahan2016droplet,KherAlden2020}. A nanobeam cantilever functionalized with a dielectric nanoparticle positioned tangentially to the droplet’s equator, oscillating in shear-force configuration, can then supply the relevant mechanical degrees of freedom. 

Beyond spherical geometries, similar symmetry-breaking couplings can also emerge in Fabry--Perot cavity architectures. In particular, a mirror capable of both axial displacement and lateral motion (or rotation about its vertex) naturally produces the two mechanical coordinates of our model. Axial motion yields the familiar dispersive coupling through modulation of the cavity length, while lateral or rotational motion breaks the transverse symmetry of the resonator and mixes the $TEM_{00}$ and $TEM_{01}$ modes, giving rise to the transverse intermode interaction described by Eq.~\ref{eq:Hamiltonian-3-mode}. Unlike the effectively one-dimensional geometry considered in Ref.~\cite{Cheung2011}, where the membrane partitions the resonator into two coupled optical subsystems, the present configuration explicitly separates symmetry-preserving and symmetry-breaking mechanical degrees of freedom, allowing frequency shifts and intermode coupling strengths to emerge as independent dynamical variables.

In Sec.~\ref{subsec:experiment} we provide a detailed feasibility analysis for the WGM setup, while analysis of other potential realizations will be presented in separate publications.

In the frame rotating with frequency $\omega_0$ the Hamiltonian of this model can be written as \begin{equation}\label{eq:Hamiltonian-3-mode}
    \begin{aligned}
        \hat{H}=&-\hbar\Delta \left(\hat{a}_{1}^{\dagger}\hat{a}_{1}+ \hat{a}_{-1}^{\dagger}\hat{a}_{-1}\right)+\hbar\Omega\left(\hat{b}_{y}^{\dagger}\hat{b}_{y}+\hat{b}_{z}^{\dagger}\hat{b}_{z}\right)\\
&+\hbar \left(\hat{b}_{z}^{\dagger}+\hat{b}_{z}\right)\left[g_1^z\left(\hat{a}_{1}^{\dagger}\hat{a}_{1}+\hat{a}_{-1}^{\dagger}\hat{a}_{-1}\right)+g^z_0\hat{a}_0^\dagger\hat{a}_0\right]\\
    &+\frac{i}{2}\hbar g_{y}\left(\hat{b}_{y}^{\dagger}+\hat{b}_{y}\right)\left[\left(\hat{a}_{-1}^{\dagger}+\hat{a}_{1}^{\dagger}\right)\hat{a}_{0}-\hat{a}_{0}^{\dagger}\left(\hat{a}_{-1}+\hat{a}_{1}\right)\right].
    \end{aligned}
\end{equation}

The first line of this Hamiltonian introduces bare optical modes described by creation-annihilation operators $\hat{a}_{\pm 1}^\dagger,\hat{a}_{\pm 1}$ and detuning $\Delta=\omega_0-\omega_1$, as well as bare mechanical modes of an isotropic harmonic oscillator of frequency $\Omega$ and creation-annihilation operators $\hat{b}_{y,z}^\dagger,\hat{b}_{y,z}$. The second line contains the standard dispersive coupling characterized by coupling parameters $g_{0,1}^z$ for optical modes $a_0$ and $a_{\pm 1}$ respectively. In what follows we shall assume for concreteness that $\Delta>0$, and $g_1^z>g_0^z$. The third line of the Hamiltonian represents a symmetry-breaking coupling of the optical modes  characterized by parameter $g_y$. This type of interaction is fairly typical in cavity and circuit QED, where it is known as transverse or beam-splitter coupling and, in the qubit–cavity context, takes the familiar Jaynes–Cummings form \cite{Blais2021RMP,Devoret2013Science,Wallraff2004Nature,Gerry2005Book,Scully1997Book}, but it hasn't been previously treated as a primary optomechanical interaction mechanism. 

While dispersive coupling parameters $g_{0,1}^z$ are quite universally defined by Eq.~\ref{eq:disp_coupling}, the definition for the symmetry breaking coupling coefficient $g_y$ depends on the details of a system described by Hamiltonian  Eq.~\ref{eq:Hamiltonian-3-mode}. For our paradigmatic WGM-nanoparticle model the interaction parameter $g_y$ can be shown to be 
\begin{equation}\label{eq:transv_coupling}
    g_y=-Lx_{zpf}\frac{\Delta}{r_0},
\end{equation}
where $r_0$ is the distance between the oscillating particle and the center of the resonator in equilibrium and $L$ is the  harmonic degree number of the respective WGM.  A detailed derivation of the Hamiltonian in Eq.~\ref{eq:Hamiltonian-3-mode} for this configuration is provided in the Supplementary Materials.

Hamiltonians of the general form of Eq.~\ref{eq:Hamiltonian-3-mode} have been introduced previously in the context of multimode optomechanical systems with mechanically induced intermode coupling~\cite{Cheung2011,Stannigel2012PRL,SafaviNaeini2011PRL}. However, the corresponding dynamical analysis was either restricted to the treatment of the transverse interaction as a time-dependent perturbation responsible for transitions between bare optical modes~\cite{Cheung2011}, or considered only in the presence of an external optical pump after linearization around a driven steady state~\cite{Stannigel2012PRL,SafaviNaeini2011PRL}. In contrast, the present work focuses on the intrinsically nonlinear Hamiltonian dynamics generated by the transverse optomechanical interaction itself in the absence of external driving. The transverse interaction therefore plays a central role in determining both the structure and stability of the resulting hybridized optomechanical modes.

Introducing  operators 
\begin{eqnarray}
    \hat{\alpha}=\frac{1}{\sqrt{2}}\left(\hat{a}_1+\hat{a}_{-1}\right)\quad ; \quad
     \hat{\beta}=\frac{1}{\sqrt{2}}\left(\hat{a}_1-\hat{a}_{-1}\right)
\end{eqnarray}
we can present Hamiltonian Eq.~\ref{eq:Hamiltonian-3-mode} as
\begin{equation}\label{eq:Hamiltonian-dark-mode}
    \begin{aligned}
        \hat{H}=&-\hbar\Delta \left(\hat{\alpha}^{\dagger}\hat{\alpha}+ \hat{\beta}^{\dagger}\hat{\beta}\right)+\hbar\Omega\left(\hat{b}_{y}^{\dagger}\hat{b}_{y}+\hat{b}_{z}^{\dagger}\hat{b}_{z}\right)\\
&+\hbar \left(\hat{b}_{z}^{\dagger}+\hat{b}_{z}\right)\left[g_1^z\left(\hat{\alpha}^{\dagger}\hat{\alpha}+ \hat{\beta}^{\dagger}\hat{\beta}\right)+g^z_0\hat{a}_0^\dagger\hat{a}_0\right]\\
&+\frac{i}{\sqrt{2}}\hbar g_{y}\left(\hat{b}_{y}^{\dagger}+\hat{b}_{y}\right)\left(\hat{\alpha}^{\dagger}\hat{a_0}-\hat{a}_{0}^{\dagger}\hat{\alpha}\right).
    \end{aligned}
\end{equation}
The variable described by operator $\hat{\beta}$ does not appear in the symmetry breaking part of the Hamiltonian, and can therefore, be considered  a "dark" mode.  Optical dark modes are well known in $\Lambda$-type and cavity-EIT schemes in QED \cite{Fleischhauer2005RMP}, where they  arise dynamically from destructive interference between two coherent coupling pathways. In contrast, in this model such a mode is  built directly into the Hamiltonian itself providing additional flexibility in controlling the system’s dynamics. At the same time the main  results presented here do not rely on its existence.

\begin{figure}
\centering
\vspace{0\linewidth}
\includegraphics[width=0.5\linewidth]{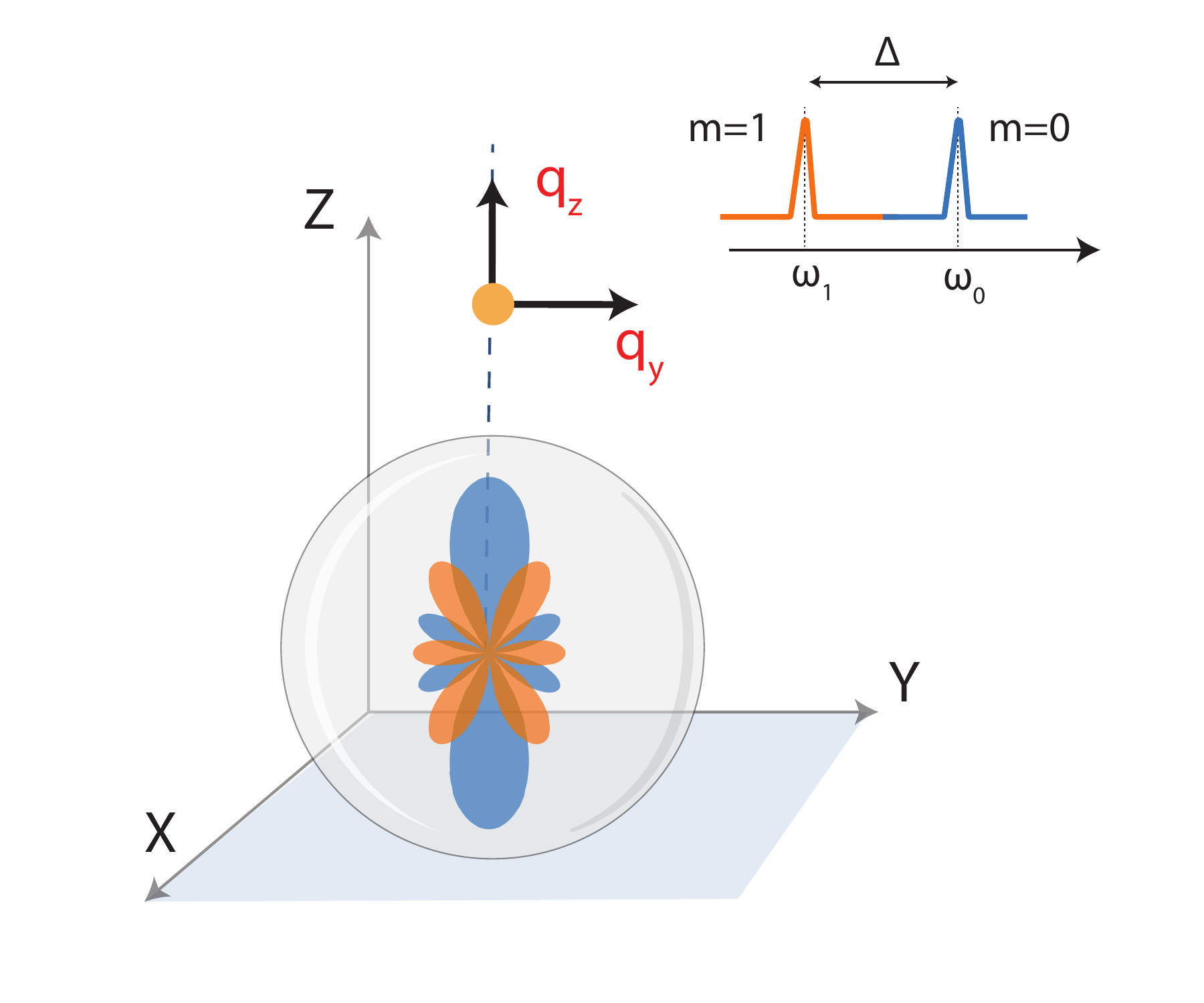}
\caption{Schematic of the model showing the $m=0$ mode (blue) with the field enhanced along the line connecting the center of the resonator with the particle and two overlapping $m=\pm 1$ modes (orange) with the field significantly reduced in this direction\cite{Deych2009,Rubin2010}. It also shows the global ($X,Y,Z$) and local coordinate axes with the components of the particle's displacement $q_y,q_z$. The inset shows the detuning between frequencies $\omega_0$ and $\omega_{\pm 1}$}
\label{fig:schematic}\end{figure}
Mechanical motion is conveniently described in terms of  dimensionless mechanical coordinates and momenta defined as:
\begin{equation}\label{eq:q_p_dimensionless}
    \hat{q}_{y,z}=\frac{1}{\sqrt{2}}\left(\hat{b}_{y,z}+\hat{b}^\dagger_{y,z}\right), \qquad 
    \hat{p}_{y,z}=i\frac{1}{\sqrt{2}}\left(\hat{b}^\dagger_{y,z}-\hat{b}_{y,z}\right),
\end{equation}
with the commutation relation 
$$[\hat{q},\hat{p}]=i.$$
In the Supplementary Materials the same notation is used for the corresponding dimensional coordinates and momenta. In the present section, however, the use of dimensionless variables is more convenient because the zero-point fluctuation amplitude  $x_{zpf}$ is already absorbed into the coupling parameters  $g^{z}_{0,1}$ and $g_y$. 

In the absence of an external pump our model has natural $SU(2)$ symmetry, which can be explicitly revealed by presenting  optical degrees of freedom in terms of pseudospin Schwinger operators \cite{Schwinger1952,Gerry2005Book} defined as 
\begin{eqnarray}\label{eq:Schwinger-operatorsJx}
 \hat{J}_{x}&	=	&\frac{1}{2}\left(\hat{\alpha}^{\dagger}\hat{a}_{0}+\hat{a}_{0}^{\dagger}\hat{\alpha}\right)\\
\hat{J}_{y}&	=&	\frac{1}{2i}\left(-\hat{\alpha}^{\dagger}\hat{a}_{0}+\hat{a}_{0}^{\dagger}\hat{\alpha}\right)\label{eq:Schwinger-operatorsJy}\\
\hat{J}_{z}&	=&	\frac{1}{2}\left(-\hat{\alpha}^{\dagger}\hat{\alpha}+\hat{a}_{0}^{\dagger}\hat{a}_{0}\right)\label{eq:Schwinger-operatorsJz}. 
\end{eqnarray}

These operators satisfy the usual commutation relations
$$
[\hat{J}_i,\hat{J}_j]=i\epsilon_{i,j,k}\hat{J}_k,
$$
where $\epsilon_{i,j,k}$ is the standard Levi-Civita symbol. The relationship between the pseudo-spin operator $\hat{J}_z$ and the photon number $\hat{N}_b$ is established by rewriting Eq.~\ref{eq:Schwinger-operatorsJz} as $\hat{J}_{z}=\hat{N}_b/2-\hat{\alpha}^\dagger\hat{\alpha}$. As in ordinary angular momentum algebra, eigenvalues of $\hat{J}_z$ take discrete values between -$N_b/2$ and $N_b/2$ (see details in the Method Section~\ref{sec:method}. 

Assuming for simplicity that $\omega_0$ does not depend on mechanical displacement so that $g_0^z=0$ and that the dark modes are not excited so that $\hat{N}_d$ can be omitted, we can rewrite Hamiltonian of Eq.\ref{eq:Hamiltonian-dark-mode}  as 
\begin{equation}\label{eq:BlochHamilttot}
    \hat{H}=-\hbar\mathbf{\hat{\Lambda}}\cdot \mathbf{\hat{J}}+\hbar\Omega\left(\hat{b}_{y}^\dagger \hat{b}_{y}+\hat{b}_{z}^\dagger \hat{b}_{z}\right)+\frac{\hbar}{\sqrt{2}}g_1^z\hat{q}_z\hat{N}. 
    \end{equation}
Here operator $\mathbf{\hat{\Lambda}}$ plays the role of an effective pseudomagnetic field and is defined as 
\begin{equation}\label{eq:magnetic_field}
    \hat{\Lambda}_x=0, \hspace{2pt}\hat{\Lambda}_y=-2g_y\hat{q}_y,\hspace{2pt} \hat{\Lambda}_z=-\left[\Delta-\sqrt{2}g_1^z\hat{q}_z\right]
\end{equation} 
(An complete form of this Hamiltonian with non-zero $g_0^z$ and $N_d$ can be found in Section~\ref{subsec:equation_motion}, Eq.~\ref{eq:Hamiltonian Schwinger}. Relaxing these assumptions leads only to a renormalization of the effective detuning parameter $\Delta$ and does not qualitatively affect the results presented below).
\section{Results}\label{sec:results}
\subsection{Linear regime and bifurcation}\label{subsec:linear}
The Hamiltonian given by Eq.~\ref{eq:Hamiltonian-dark-mode} generates dynamics with two conserving quantities: total photon number $N$
\begin{equation}
\hat{N}=\hat{\alpha}^{\dagger}\hat{\alpha}+\hat{\beta}^{\dagger}\hat{\beta}+\hat{a}_{0}^{\dagger}\hat{a}_{0}
\end{equation}
and the photon number in the dark mode $N_d$:
\begin{eqnarray}
    N_d=\hat{\beta}^\dagger\hat{\beta}
\end{eqnarray}
Obviously, the number of photons in the remaining bright modes $N_b=N-N_d$ is also conserved, and after restricting the dynamics to the bright subspace by setting  $N_d=0$ in Eq.~\ref{eq:BlochHamilttot} , we have $N_b=N$. Hamiltonian of Eq.\ref{eq:BlochHamilttot} generates Heisenberg equations for mechanical variables of the form
\begin{eqnarray}
    \frac{d\hat{q}_{z}}{dt}&	=	& \hat{p}_{z};\label{eq:qz_simpl}\\
\frac{d\hat{p}_{z}}{dt}&	=&	- \hat{q}_{z}-\frac{1}{\sqrt{2}}g_{1}^{z}\left(\hat{N}-2\hat{J}_{z}\right)\label{eq:pz_simple}\\ 
\frac{d\hat{q}_{y}}{dt}	&=	& \hat{p}_{y}\label{eq:qy_simpl}\\
\frac{d\hat{p}_{y}}{dt}&	=&	- \hat{q}_{y}-2g_{y}\hat{J}_{y},\label{eq:py_simple}
\end{eqnarray}
while dynamic equations for optical variables can be written in the form of the effective Bloch equations  with $\mathbf{\Lambda}$ defined in Eq.~\ref{eq:magnetic_field}
\begin{equation}\label{eq:Bloch_simpl}
    \frac{d \hat{\mathbf{J}}}{dt}=\hat{\mathbf{J}}\times \mathbf{\hat{\Lambda}}
\end{equation}
In these equations all frequency-dimensional quantities are normalized by the mechanical frequency $\Omega$, while time is normalized by $1/\Omega$. Therefore, from this point forward we set $\Omega=1$. For notational simplicity, we again retain the same symbols $\Delta$ and $g_{y,z}$ for both the original (dimensional) and the corresponding dimensionless parameters. The distinction between dimensional and dimensionless quantities will be clear from context.

From this point onward we transition to a fully classical description by replacing all operators with their expectation values and neglecting quantum fluctuations. Formally, this manifests itself in dropping the $\hat{\phantom{x}}$ decorations from all dynamical variables.

We begin by describing the structure of the fixed points characterizing the dynamics generated by  Eq.~\ref{eq:qz_simpl}-\ref{eq:py_simple} and Eq.~\ref{eq:Bloch_simpl}.  The two equilibrium states of the system then can be described in terms of the original optical amplitudes $a_0$ and $\alpha$ as 
\begin{equation}\label{eq:unstable_eq}
\begin{split}
    &q_y=p_y=p_z=q_z=0\\
    &\alpha=0,\hspace{2pt}a_0=\sqrt{N} 
\end{split}
\end{equation}
and 
\begin{equation}\label{eq:stable_eq}
\begin{split}
    q_y=p_y=&p_z=0\quad;\quad q_z=-\sqrt{2}g_1^zN\\
    &\alpha=\sqrt{N};\hspace{2pt}a_0=0 
\end{split}
    \end{equation}
The more general expressions for the fixed points without the restrictions on $g_0^z$ and $N_d$ are given in Section~\ref{sec:method}. 

The difference between the two fixed points becomes clear if we rewrite them in terms of Schwinger variables, where the optical component of the fixed-point manifold is characterized by $J_z=\pm N/2$. The difference between these two configurations lies in their stability: $J_z=N/2$ corresponds to an unstable optical equilibrium, while the state with $J_z=-N/2$ represents a stable equilibrium of the optical subsystem (see Eq.~\ref{eq:BlochHamilt}).

The dynamics near optically stable and unstable equilibria are fundamentally different. In this work we focus on the latter, where an interesting phenomenon emerges: mechanical motion stabilizes the optical dynamics until the optical excitation strength, quantified by the photon number $N$, exceeds a critical value resulting in a Hamiltonian Hopf bifurcation. The dynamics around the optically stable equilibrium will be addressed in a separate paper.

Linearizing   Eq.~\ref{eq:qz_simpl}-\ref{eq:Bloch_simpl} around fixed point defined by Eq.~\ref{eq:unstable_eq}  we find that linear optomechanical oscillations of coupled $q_y$ and $\alpha$ variables are characterized by  the following normal frequencies (see details in the Method section):
\begin{equation}\label{eq:eigenfreq}
\begin{split}
    \lambda_{1,2}&=\pm \sqrt{\frac{(\Delta^{2}+1)+(\Delta^{2}-1)\sqrt{1-N/N_c}}{2}}\\
    \nu_{1,2}&=\pm \sqrt{\frac{(\Delta^{2}+1)-(\Delta^{2}-1)\sqrt{1-N/N_c}}{2}}
\end{split}
\end{equation}
where $N_c$ is 
\begin{equation}
N_{c}=\frac{(\Delta^{2}-1)^2}{8 g_y^{2}\Delta}
\label{eq:Nc}\end{equation}
When the photon number  $N$ exceeds the critical value $N_c$ the eigen-frequencies split into complex-conjugate pairs with nonzero imaginary parts, and the system undergoes a Hamiltonian Hopf bifurcation (also known as Krein collisions)\cite{Haller1991}. Physically, this instability develops around an optically unstable but mechanically stabilized equilibrium, when the stabilizing effect of the mechanical motion can no longer compensate for the intrinsic optical instability.

At $\Delta=1$ (when the detuning between the optical modes is in exact resonance with the mechanical frequency), the mechanical stabilization mechanism collapses because the critical photon number $N_c$ vanishes. Physically, this means that the optically unstable equilibrium cannot be stabilized even at arbitrarily small optical excitation. Interestingly, the same resonance condition plays a central role in the perturbative transition dynamics considered in Ref.~\cite{Cheung2011}, where it corresponds to the maximization of the transition probability. In the present work, however, it emerges instead as the singular point separating mechanically stabilizable and non-stabilizable Hamiltonian regimes.

It is important to emphasize that the Hamiltonian Hopf bifurcation  is different from the Hopf bifurcations in dissipative systems, which result in formation of limit cycles familiar from self-sustained optomechanical oscillations\cite{Marquardt2006,Ludwig2008}.  As expected for the Hamiltonian Hopf bifurcation, the loss of stability of the  fixed point does not lead to the emergence of a new equilibrium or a limit cycle. Instead, the motion remains confined to invariant oscillatory manifolds determined by the conserved quantities whose values are fixed by the initial conditions. 

The numerical value of $N_c$ depends on the experimental realization of the model, but estimates performed for a droplet spherical WGM resonator indicates that there exists an experimentally realizable set of parameters for which this transition might occur at relatively low photon numbers corresponding to the intracavity optical powers in the $\mu W$ region (see details in the Experimental feasibility  section \ref{subsec:experiment}). 

\subsection{Nonlinear dynamics: Simulation}\label{subsec:nonlinear}
Numerical simulation of the dynamics of the system are carried out with initial conditions of the following form:
\begin{equation}\label{eq:initial_condition}
\begin{split}
p_y=q_z=p_z=&\alpha=\beta=0\\
q_y(0)=q_0, &\hspace {5 pt} a_0=\sqrt{N}
\end{split}
\end{equation}
where $N$ is the total number of photons in the system. These initial conditions correspond to the dynamics around optically unstable equilibrium, and are used throughout the rest of this paper.  Since  the dark optical mode is not excited $N_b=N$. 
The two parameters $q_0$ and $N$ define the dynamical properties of the system with the former controlling the degree of nonlinearity, and the latter the dynamic regime.

The system of equations presented in Eq.~\ref{eq:qz_simpl} - \ref{eq:Bloch_simpl} is numerically solved with the specified initial conditions. We monitor the mechanical ($E_{mech}$) and optical ($E_{opt}$) energies in both the normalized time ($\Omega t$) and frequency ($\omega$) domains, for various degrees of nonlinearity, $q_0$ and the initial photon numbers $N$. As a check for the numerical accuracy we verify that the total energy of the system (including the interaction energy) conserves in the course of simulations.  

The system dynamics depends sensitively on the parameters $g_y$ and $g_{1}^z$. In selecting their numerical values, our goal was to keep the critical photon number $N_c$ in the range $10^6$–$10^7$, thereby demonstrating the feasibility of reaching the bifurcation at low optical powers. This requirement imposes stringent constraints on the detuning $\Delta$. On one hand, the resonance condition $\Delta=1$ corresponds to the collapse of the mechanical stabilization mechanism discussed above. On the other hand, moving too far away from resonance rapidly increases the critical photon number $N_c$. As a result, experimentally interesting regimes require detunings sufficiently close to resonance while remaining outside the singular point $\Delta=1$. In a WGM realization of the model, this constraint leads to relatively small values of the coupling parameters $g_y$ and $g_{1}^z$.

Using the estimates provided in Sec.~\ref{subsec:experiment}, we set 
\begin{equation}\label{eq:parameters}
g_y = -3\times10^{-5}, \hskip 2pt g_1^z = -4.2\times 10^{-5}, \hskip 2pt \Delta = 1.1.
\end{equation}
For these parameters, the critical photon number is $N_c\simeq 10^7$, corresponding to an intracavity power of $P\sim 0.1 mW$. The  initial displacement $q_0$ is chosen in the range  $10^3$ (weakly nonlinear regime) to $3\times 10^5$ (strongly nonlinear regime). To put these values in perspective, the relevant spatial scale is set by the zero-point fluctuation amplitude $x_{zpf}$. Using the estimates of Sec.~\ref{subsec:experiment}, we find that  $q_y=10^3$ corresponds to physical displacement of about $1.6~\text{nm}$, which is comparable to thermal fluctuations at room temperature. In contrast, $q_0\sim 3\times 10^5$ corresponds to larger displacements of the order of $0.3~\mu m$, but still within the linear limit of nano-beam cantilevers. 

\subsubsection{Stable Regime}\label{subsubsec:Stable_regime}
\begin{figure}
    \centering
     \begin{subfigure}{\textwidth}
         \centering     \includegraphics[width=\textwidth]{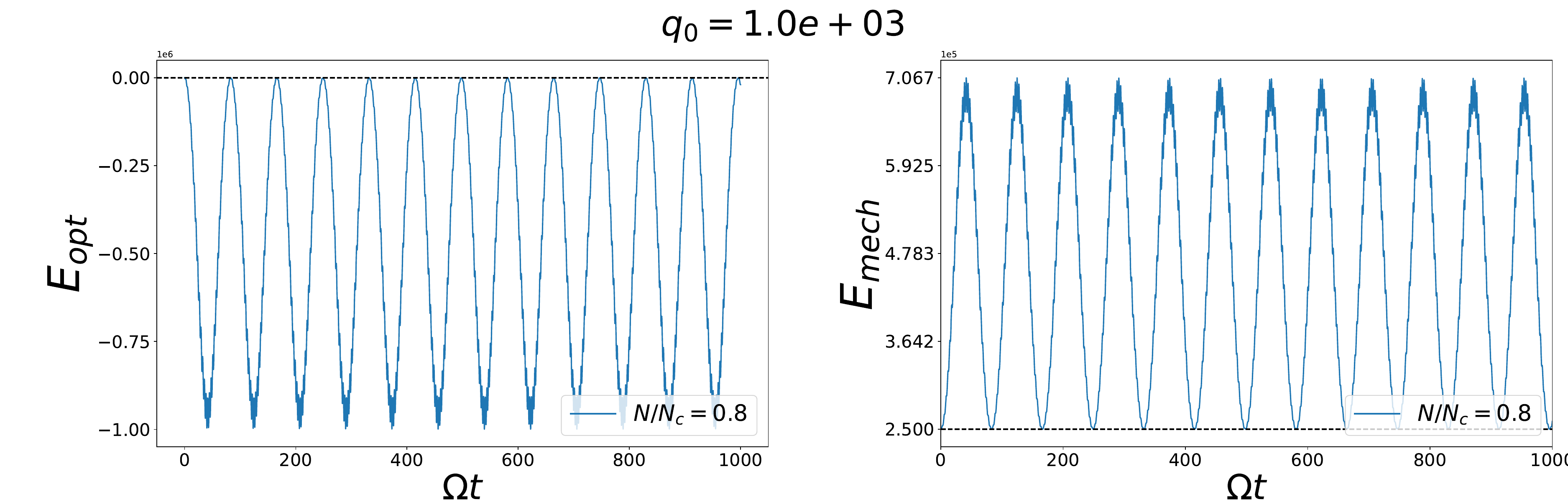}
         \end{subfigure}
         \vfill
     \begin{subfigure}{\textwidth}
         \centering         \includegraphics[width=\textwidth]{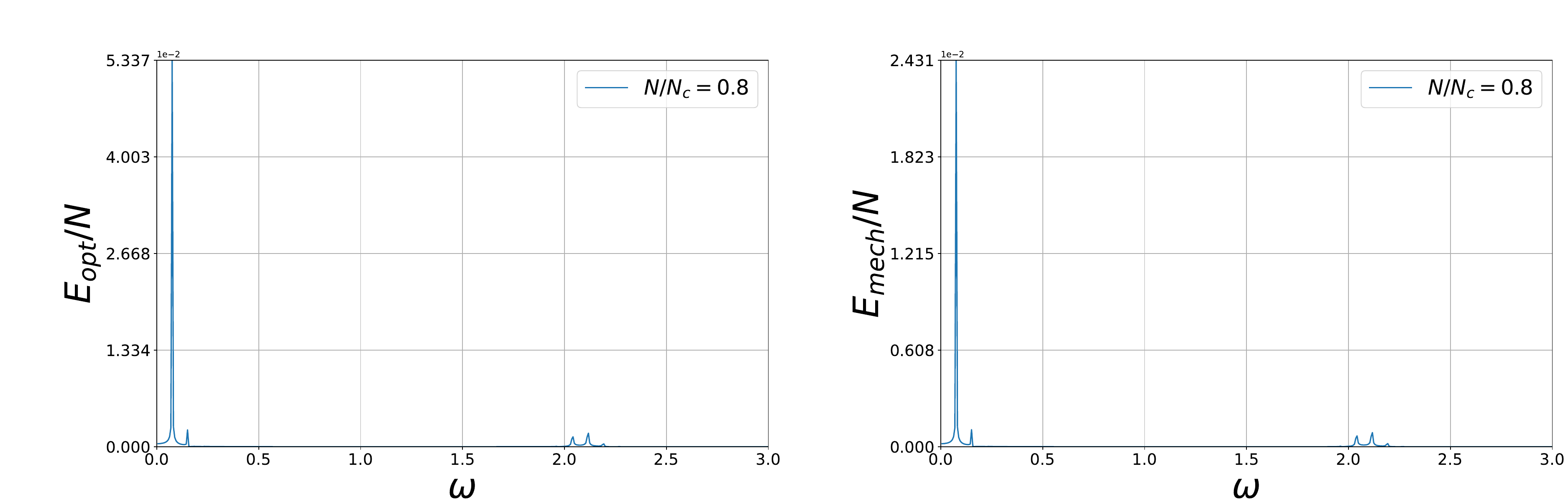}
   \end{subfigure}   
     \caption{Evolution of mechanical and optical energies (top) and the corresponding Fourier spectra (bottom) for $N=0.8N_c$ for weakly non-linear case $q_0=10^3$. The time-dependence and respective spectra reflect quadratic dependence of the energy on mechanical amplitudes displaying oscillations at frequencies $2\lambda$, $2\nu$, and $\lambda\pm\nu$. The last plot in the top row indicates that in the case of weak nonlinearity the mechanical energy of the system always exceed its initial value, presented by the horizontal line so that the energy flows from optical to mechanical degrees of freedom.   }
\label{fig:stable-evolution-low}
\end{figure} 
For relatively small nonlinearity, the dynamics is dominated by oscillations at the linear normal frequencies, with amplitudes modulated by the nonlinearity. However, for the chosen values of $\Delta$, $\Delta-1\ll 1$ the difference between normal frequencies $\lambda-\nu$ is of the order of 
\begin{equation}\label{eq:freq-diff}
\lambda-\nu\approx \left(\Delta-1\right)\sqrt{1-N/N_c} \ll 1
\end{equation}
making it difficult to distinguish between $\lambda+\nu$, $2\lambda$ and $2\nu$ spectral features. This difference becomes even smaller when $N$ approaches the critical value $N_c$. For this reason, the plots in Fig.\ref{fig:stable-evolution-low} demonstrate a single prominent spectral feature at the low frequency $\lambda-\nu$ and much weaker features around three other frequencies.  

For completeness we also provide results of the simulations in this regime for a much bigger value of $\Delta=10 $ corresponding to much softer mechanical frequency, which also increases dimensionless values of parameters $g_{y,z}$. Such values can be realized in the WGM-dipole setup if we discard the idea of observing the bifurcation transition, which in this case would require much larger optical powers. As an illustration we provide the results of calculations with $g_y=0.1$ and $q_0=1$ in Fig. \ref{fig:stable-evolution-low_old} depicting the time dependence of the dimensionless optical ($E_{opt}$) and mechanical energies ($E_{mech}$) together with their Fourier spectra. Here we have a much richer spectrum, which shows the peaks at all four distinct frequencies $2\lambda$, $2\nu$ and $\lambda \pm\nu$ as well as their evolution with increasing number of photons $N$.  The temporal traces reveal a clear modulation pattern with  the modulation frequency, which can be found analytically after averaging out the fast oscillations:  (see details in the Supplementary Materials):
\begin{figure}
    \centering
     \begin{subfigure}{\textwidth}
         \centering     \includegraphics[width=\textwidth]{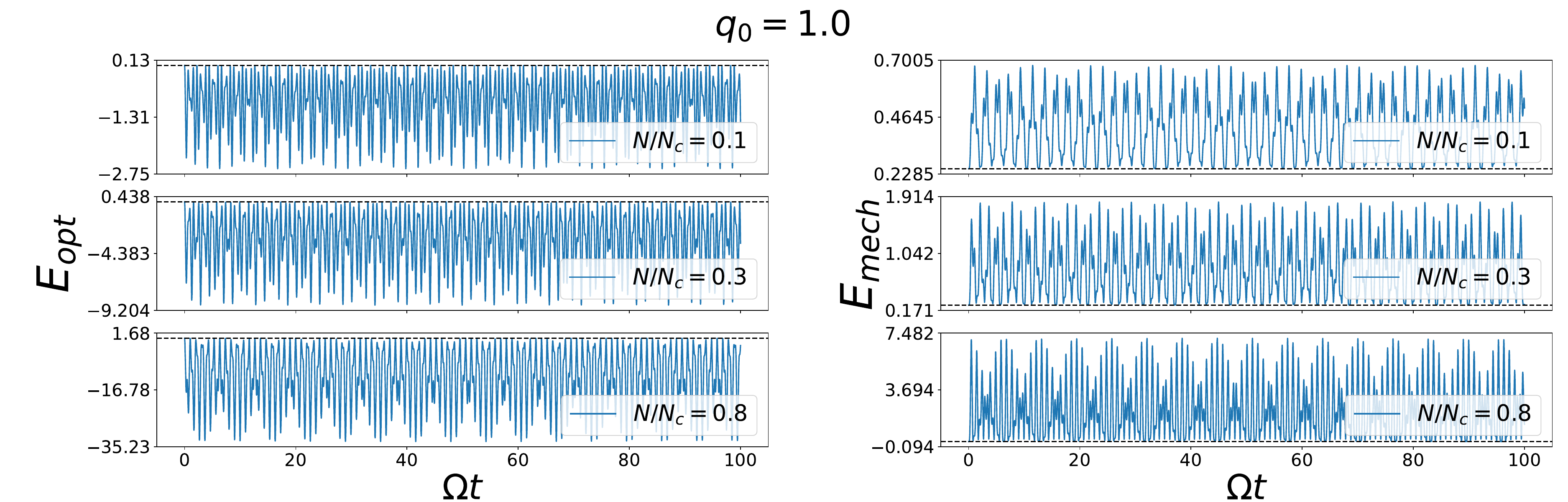}
         \end{subfigure}
         \vfill
     \begin{subfigure}{\textwidth}
         \centering         \includegraphics[width=\textwidth]{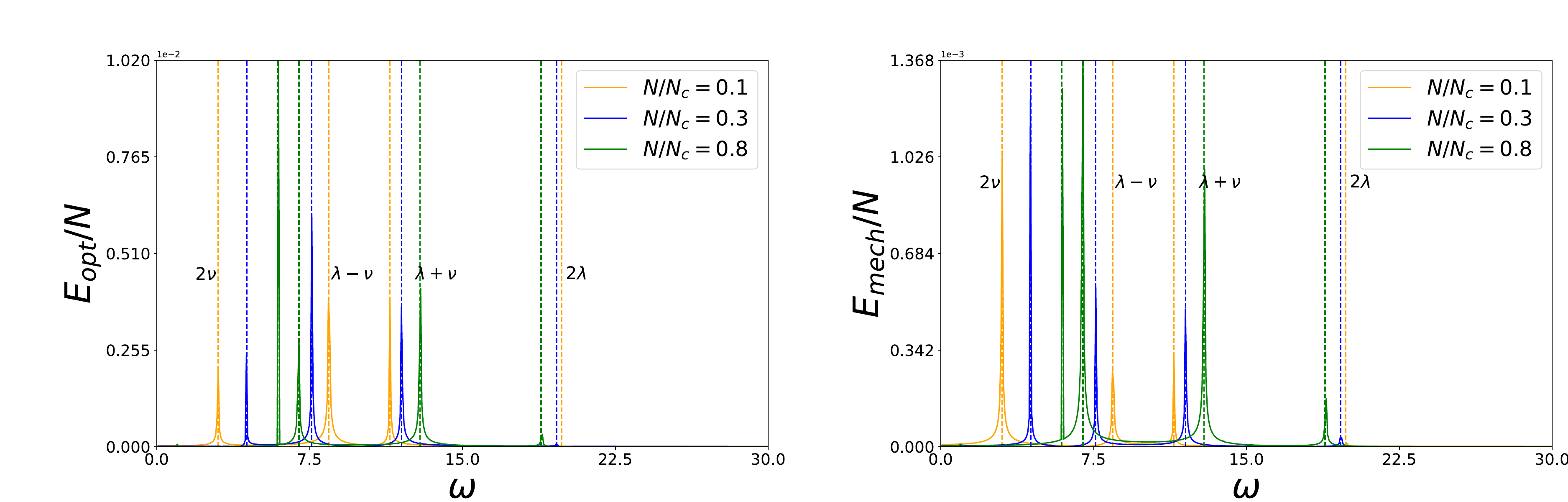}
   \end{subfigure}   
     \caption{Evolution of mechanical and optical energies (top) and the corresponding Fourier spectra (bottom) for various photon numbers $N$ for $\Delta=10$, $g_y\approx 0.3$ and $q_0=1$. The time-dependence and respective spectra reflect quadratic dependence of the energy on mechanical amplitudes displaying oscillations at frequencies $2\lambda$, $2\nu$, and $\lambda\pm\nu$. The last plot in the top row indicates that in the case of weak nonlinearity the mechanical energy of the system always exceed its initial value presented by the horizontal line, so that the energy flows from optical to mechanical degrees of freedom.   }
\label{fig:stable-evolution-low_old}
\end{figure} 

\begin{equation}\label{eq:modulation_freq}
 \omega_{mod}=\frac{\left|g_y\right| q_0}{2\sqrt{2}},
\end{equation}

\begin{figure}
\centering
\begin{subfigure}{\textwidth}
         \centering \includegraphics[width=\textwidth]{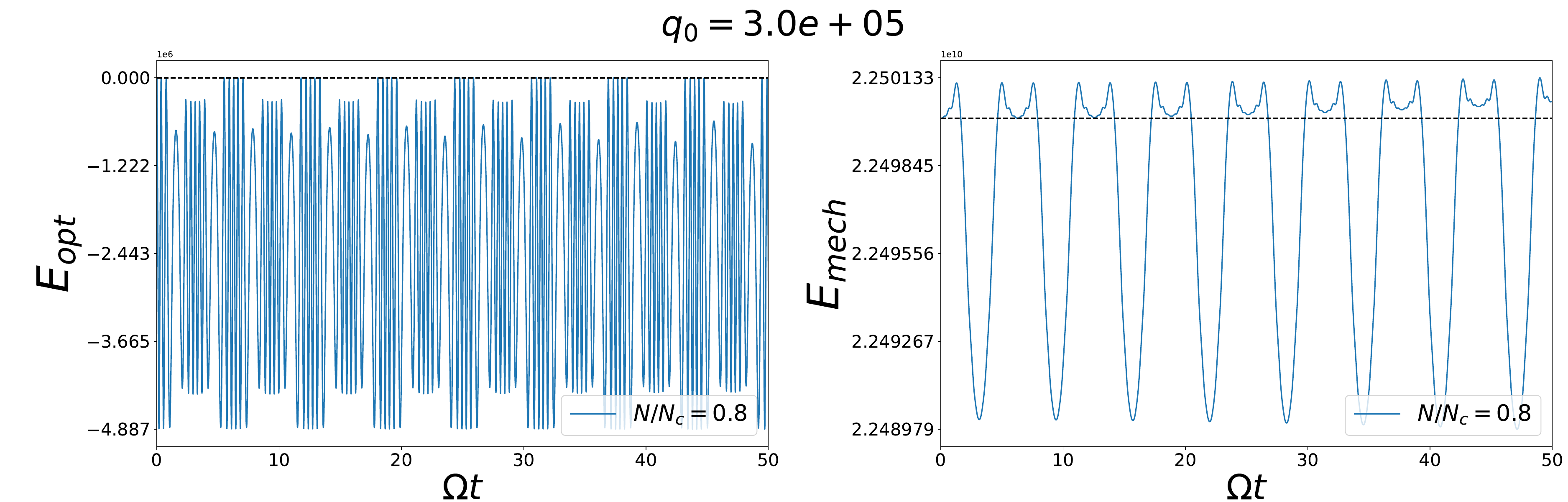}
         \end{subfigure}
         \vfill
     \begin{subfigure}{\textwidth}
         \centering        \includegraphics[width=\textwidth]{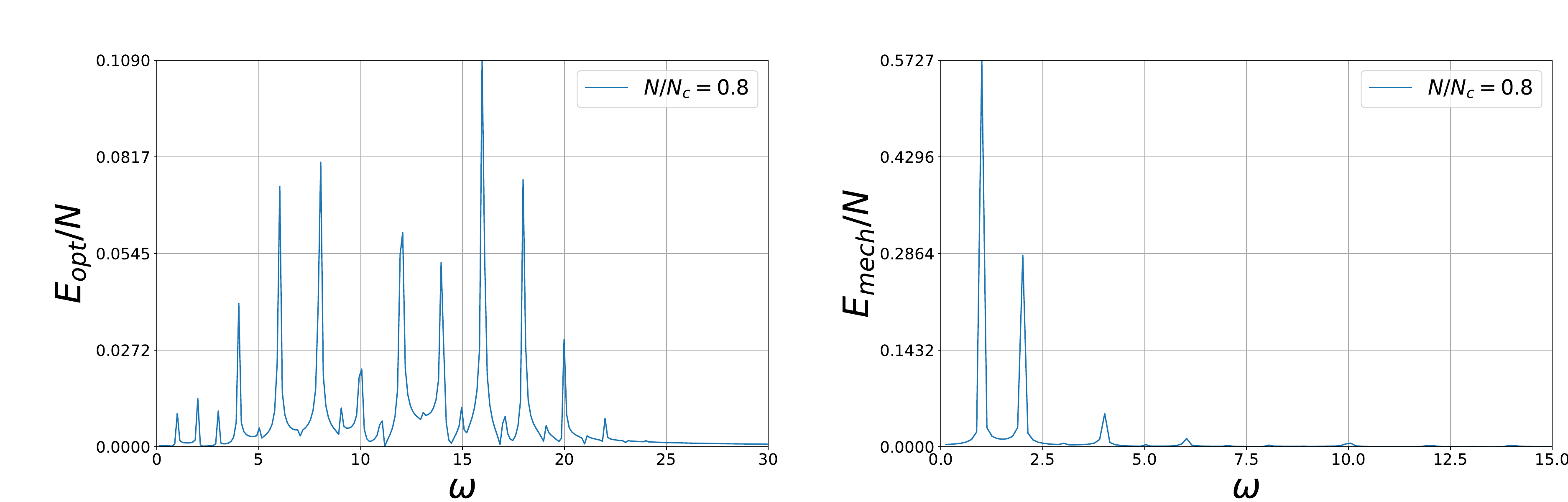}
     \end{subfigure}
\caption{Evolution of mechanical and optical energies (top) and the corresponding Fourier spectra for various photon numbers $N$ for strong non-linearity  $q_0=3\times10^5$. The comparison of mechanical energy with its initial value presented by a horizontal line in the last plot of the top row indicates that in this case the mechanical energy drops below its initial value in pulses indicating the flow of energy from mechanical to optical degrees of freedom. The large offset of the mechanical energy is the result of the normalization used in Eq.\ref{eq:ene-mech}.}\label{fig:stable-evolution-high}
\end{figure}

  This picture breaks down when nonlinearity becomes stronger as illustrated by  Fig.\ref{fig:stable-evolution-high}. (Here we are reverting back to the values of the parameters given in Eq.~\ref{eq:parameters}). Analytical treatment in this case is not possible as time evolution becomes more complex and is characterized by multiple sidebands.  A notable feature of this result is that while the mechanical spectrum exhibits strong sidebands within a relatively narrow interval of order $3\Omega$, the optical spectrum develops a broad distribution of higher-order sidebands spanning a much wider range, of order $18\Omega$, with the higher-order sidebands carrying spectral weight comparable to that of the first low frequency lines. Although the intrinsic radiation-pressure nonlinearity can generate higher-order sidebands at multiples of the mechanical frequency in dispersive optomechanical systems, their amplitudes are strongly suppressed in the weak-modulation regime typical of standard optomechanics\cite{Aspelmeyer2014,Schliesser2008}. In contrast, the transverse coupling generates an extended multi-sideband structure already at moderate initial displacements. It shall be noted that even low frequency peaks do not correspond to the standard Stokes and anti-Stokes lines because they do not origin from mechanical modulation of optical frequencies. The origin of these lines is in the mechanical modulation of the intermode coupling with its own characteristic frequency $\Delta$, which itself is comparable to the mechanical frequency. This proximity of the frequencies  explains a much stronger nonlinear response in this case. 

  As discussed in the Introduction, a detailed treatment of dissipation lies outside the scope of the present work. Nevertheless, in order to verify the robustness of the predicted spectral features, we performed numerical simulations including standard optical and mechanical dissipation terms in the dynamical equations. The results, presented in the Supplementary Materials, demonstrate that the extended optical sideband spectrum survives even when the optical dissipation rate becomes comparable to the mechanical frequency. As expected in this regime, dissipation suppresses primarily the low-frequency sidebands associated with long-time dynamics, while preserving the higher-frequency spectral components originating from shorter-time optical oscillations.
  
 To conclude this subsection, let us note that the optical energy presented in Fig.\ref{fig:stable-evolution-low}---\ref{fig:stable-evolution-high} never exceeds its initial value, which in our rotating frame is zero.  This limitation is obvious from expression for an optical energy, which, in the case $N_d=0$ is reduced to
 \begin{equation}
E_{opt}=-\hbar\Delta\left|\alpha\right|^2<0
 \end{equation}
The behavior of the mechanical energy, on the other hand, depends on the degree of nonlinearity. At low nonlinearity it never decreases below its initial value, so that the energy flows from optical to mechanical degrees of freedom. At stronger nonlinearity, however, the situation changes, and as one can see from Fig.\ref{fig:stable-evolution-high} the energy starts flowing in both directions - to and from the mechanical degrees of freedom. 

\subsubsection{Unstable Regime}\label{subsubsec:unstable-regime}
Since the system under consideration is Hamiltonian, its nonlinear dynamics does not exhibit asymptotically stable attractors. Consequently, a Hopf bifurcation does not generally lead to exponential growth followed by saturation at a steady-state amplitude, as is typical in dissipative systems such as lasers~\cite{Siegman1986} or optomechanical systems with parametric instability~\cite{Braginsky2001,Marquardt2006,Ludwig2008}. Instead, the dynamics in the linearly unstable regime is constrained by conservation of energy and total photon number, depends sensitively on the initial conditions, and exhibits nonlinear arrest of the instability from the onset.

\begin{figure}
\centering
\begin{subfigure}{\textwidth}
         \centering \includegraphics[width=\textwidth]{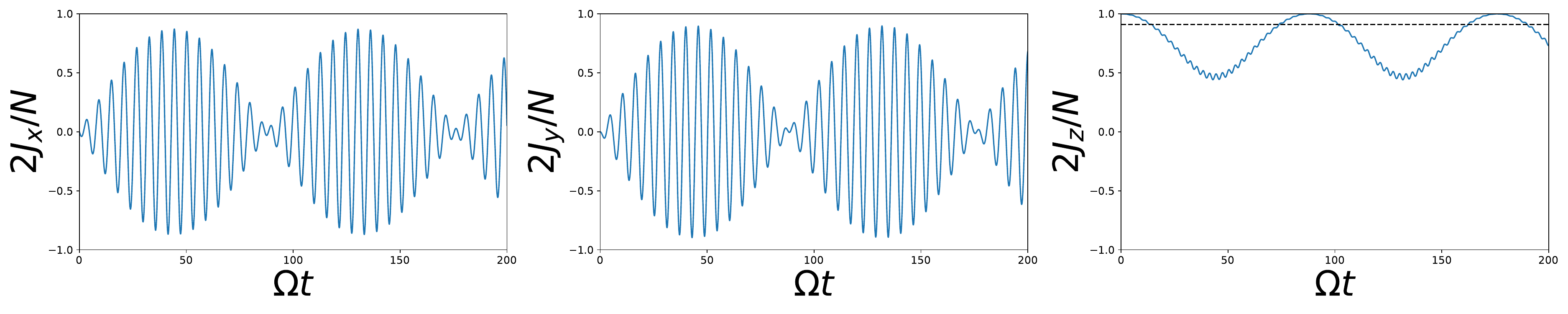}
\end{subfigure}
\vfill
\begin{subfigure}{\textwidth}
         \centering \includegraphics[width=\textwidth]{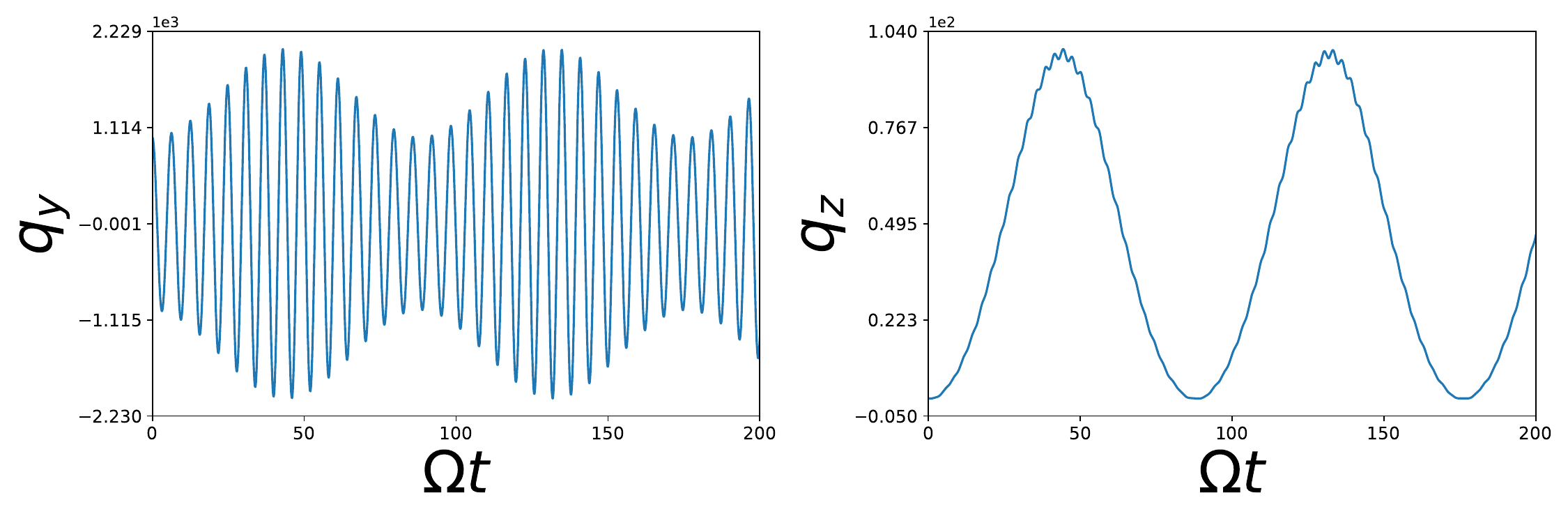}
\end{subfigure}
\caption{Time evolution of optical ($J_x,J_y,J_z$) and mechanical ($q_y,q_z$) variables in the unstable regime for a weakly nonlinear initial condition with $q_0=10^3$ and $N/N_c=1.1$. The plots show a clear separation of time scales: $J_{x,y}$ and $q_y$ exhibit amplitude-modulated fast oscillations with frequencies close to the quasi-linear normal frequencies with adjusted value of $J_z$, while $J_z$ and $q_z$ display slow oscillations at a lower frequency determined by the time-averaged self-consistent nonlinear dynamics of the system. The horizontal line on the $J_z$ plot indicates the critical value $N_c/N$, illustrating the main nonlinear stabilization mechanism. }
\label{fig:oscillations_unstable_lowN}
\end{figure}

Fig.\ref{fig:oscillations_unstable_lowN} presents the results of numerical simulations at the same low nonlinearity value of $q_0=10^3$ as in Fig.\ref{fig:stable-evolution-low} but slightly above the bifurcation threshold at $N=1.1N_c$. Instead of energy, we plot here the dynamical variables $q_{y,z}$ and $J_{x,y,z}$, because they now exhibit qualitatively different  behavior. All components of vector $\mathbf{J}$ are normalized by the equilibrium value of $J_z^{eq}=N/2$. The dynamics of these quantities differs qualitatively from the regime $N<N_c$. This behavior can be understood in terms of a self-consistent stabilizing feedback mechanism in which the system is dynamically driven across the Hopf boundary. The key observation is that the instantaneous normal-mode frequencies of $(q_y,p_y,J_x,J_y)$ subsystem become determined by the instantaneous rather than the  equilibrium value of $J_z$.  For $J_z=N/2$  and $N>N_c$, the corresponding normal frequencies acquire an imaginary part, indicating instability. However, for sufficiently large deviations of $J_z$ from equilibrium, the same normal frequencies can become purely real again, corresponding to a temporary adiabatic return of the system to the linearly stable regime. This dynamic renormalization of the instantaneous spectrum provides the mechanism by which nonlinear effects arrest the instability and stabilize the motion.
\begin{figure}
\centering
\begin{subfigure}{\textwidth}
         \centering \includegraphics[width=\textwidth]{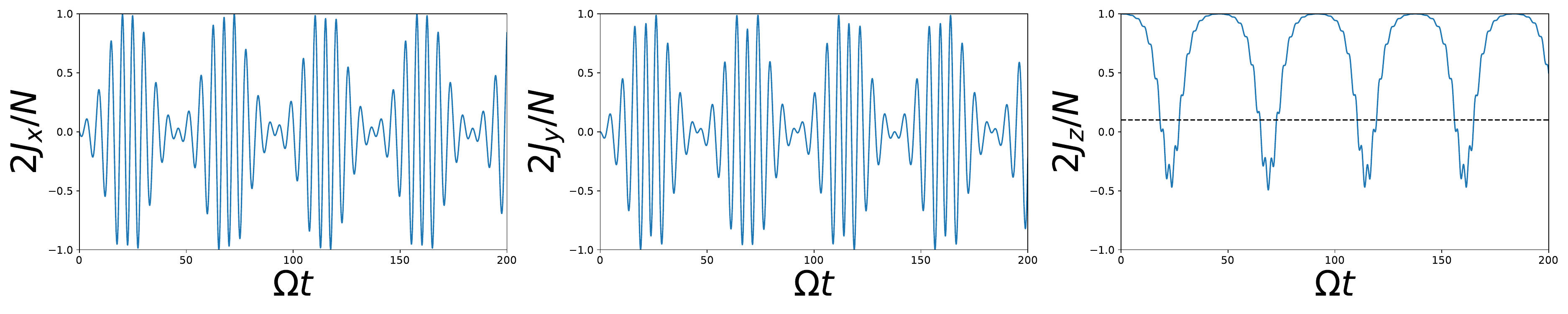}
\end{subfigure}
\vfill
\begin{subfigure}{\textwidth}
         \centering \includegraphics[width=\textwidth]{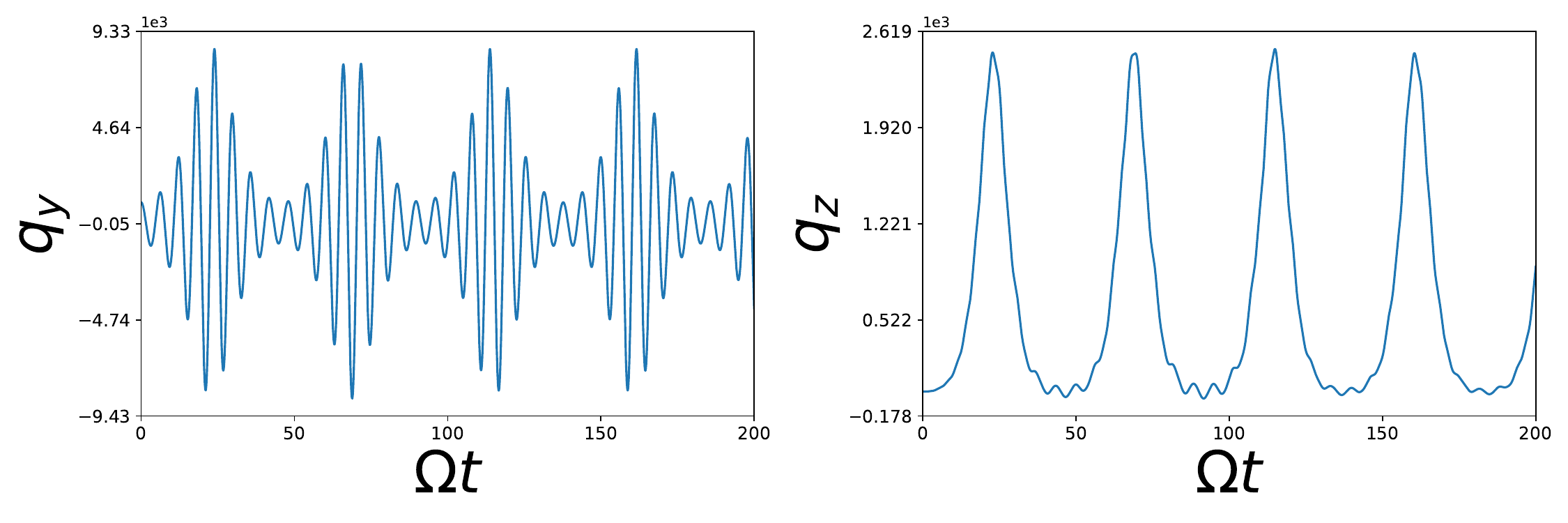}
\end{subfigure}
\caption{Time evolution of optical ($J_x,J_y,J_z$) and mechanical ($q_y,q_z$) variables in the unstable regime for a weakly nonlinear initial condition with $q_0=10^3$ and $N/N_c=10$. Compared to the case $N=1.1N_c$, the dynamics of the former slow variables becomes strongly anharmonic and the adiabatic approximation breaks down. Nevertheless, the main stabilization mechanism remains the same, as seen from the intersections of the $J_z$ trajectory with the $N_c/N$ line.}
\label{fig:oscillations_unstable_highN}
\end{figure}
This picture can be formulated more precisely in terms of quasi-linear modes and an adiabatic treatment of the slowly varying variable $J_z$, made possible by the clear separation of time scales visible in Fig.~\ref{fig:oscillations_unstable_lowN}.The variables $(q_y,p_y,J_x,J_y)$ evolve predominantly on the fast time scale set by the instantaneous normal frequencies of the linearized subsystem, while $J_z$ evolves according to Eq.~(\ref{eq:Bloch}),
\begin{equation}\label{eq:JZeq}
    \dot J_z = -2 g_y q_y J_x .
\end{equation}
where the product of the two fast-oscillating quantities $q_y$ and $J_x$ generates both fast $(\lambda+\nu)$ and slow $(\lambda-\nu)$ forcing components. At first sight one might expect that as the sum-frequency term averages out, the difference-frequency component would drive $J_z$. However, because the amplitudes of both $q_y$ and $J_x$ depend on the instantaneous value of $J_z$, Eq.\ref{eq:JZeq} does not describe a simple system driven by a periodic force. Instead, Eq.~\ref{eq:JZeq} becomes a nonlinear equation for $J_z$ with time-dependent coefficients and intrinsic dynamics of its own. A detailed analysis of this equation would take us away from the main focus of the paper. We therefore only note that the range of variation of $J_z$ observed in Fig.~\ref{fig:oscillations_unstable_lowN} is relatively small, which allows Eq.~(\ref{eq:JZeq}) to be linearized around an average value $\bar J_z$. The resulting dynamics corresponds to numerically observed harmonic oscillations of $J_z$. 

Nevertheless, the overall interpretation of the dynamics in terms of a separation of time scales remains valid. The variable $J_z$ can still be treated as quasistatic in the adiabatic approximation, while the fast variables evolve as quasi-linear modes evaluated at the slowly varying  instantaneous value of $J_z$ or its average value $\bar J_z$. As the instability develops, the growth of the fast variables drives a slow evolution of $J_z$, which feeds back into the fast subsystem through the term $2g_yJ_z q_y$ in the equation for $J_x$. This feedback dynamically renormalizes the instantaneous spectrum and can drive the system across the Hopf boundary, restoring real normal frequencies and arresting the instability. The slow component of $J_z$ provides the dominant part of this feedback, while its small fast component produces only rapidly oscillating corrections that average out to leading order. The resulting bounded motion is therefore governed by a nonlinear Hamiltonian self-stabilization mechanism.

This simple picture breaks down as the system is driven further beyond the bifurcation threshold. Fig.~\ref{fig:oscillations_unstable_highN} shows the dynamics of all oscillating variables for the same value of $q_0$, but with $N=10N_c$. While the oscillations of $q_y$ and $J_{x,y}$ can still be interpreted as fast motion modulated by a slower amplitude variation, their frequencies no longer correspond to the quasi-linear frequencies $\lambda$ and $\nu$. At the same time, the dynamics of $q_z$ and $J_z$ becomes strongly anharmonic, involving multiple frequency components and no longer admitting an adiabatic description. Nevertheless, the underlying stabilization mechanism remains the same: oscillations of $q_y$ and $J_x$ dynamically reduce the value of $J_z$ below the instantaneous bifurcation threshold and may even reverse its sign, thereby transforming the dynamics linearized around the instantaneous value of $J_z$ from unstable to stable. 

\begin{figure}
\centering
\begin{subfigure}{\textwidth}
         \centering \includegraphics[width=\textwidth]{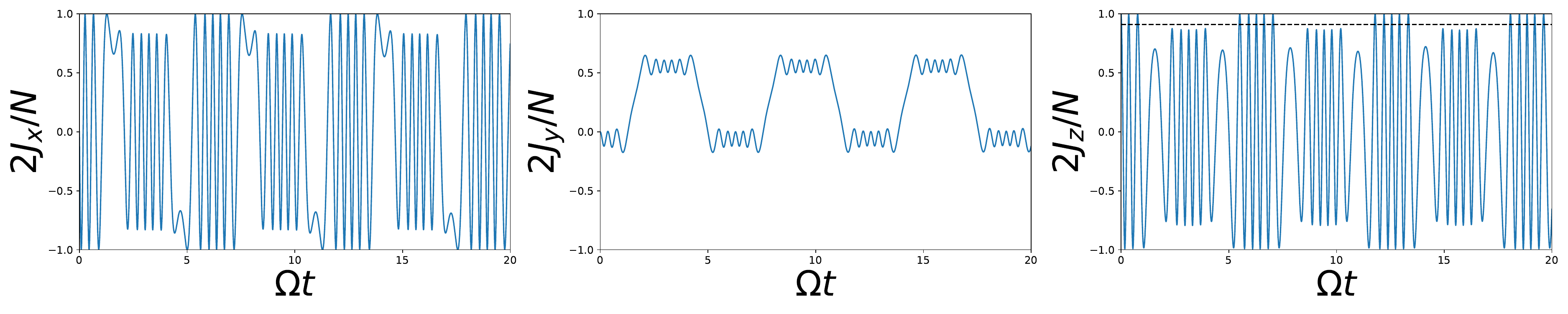}
\end{subfigure}
\vfill
\begin{subfigure}{\textwidth}
         \centering \includegraphics[width=\textwidth]{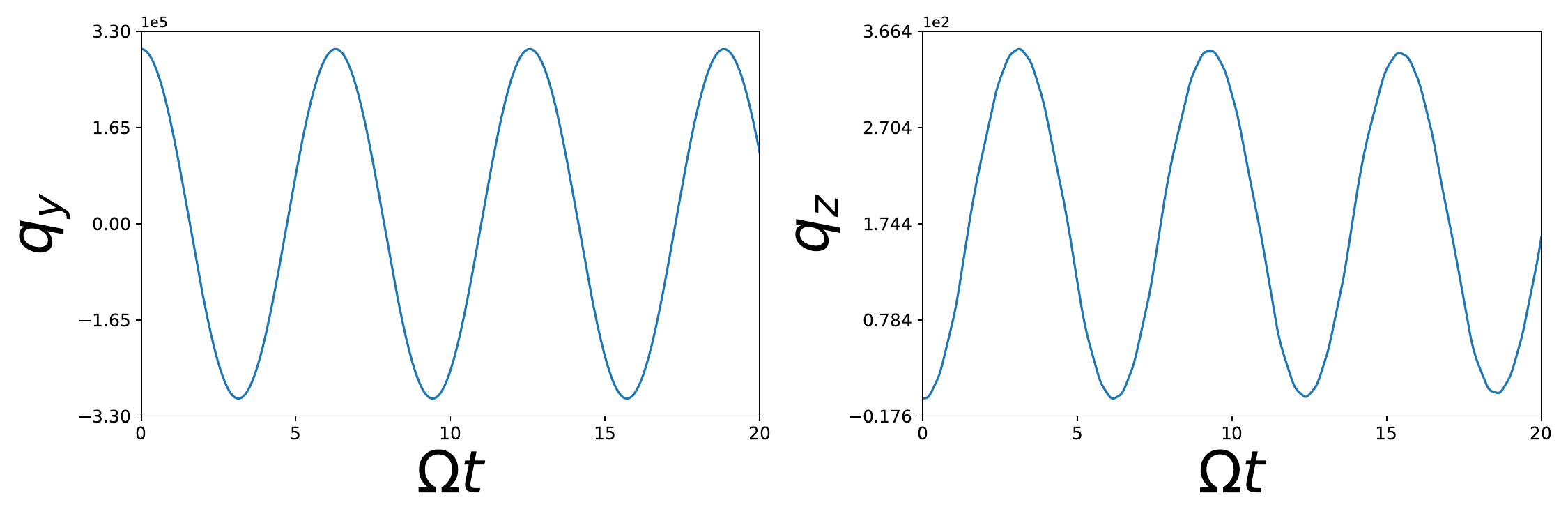}
\end{subfigure}
\caption{Evolution of optical ($J_x,J_y,J_z$) and mechanical ($q_y,q_z$) variables in the unstable regime corresponding to strongly non-linear case with $q_0=3\times10^5$ and $N/N_c=1.1$. Nonadiabatic disruptions in the optical modes coincide with mechanical variables vanishing.}
\label{fig:unstable-large-q}
\end{figure}

As we move toward stronger nonlinearity with $q_0=3\times10^5$ and $N=1.1N_c$, the system dynamics undergoes a qualitative change (see Fig.~\ref{fig:unstable-large-q}). The striking feature of this regime is the apparent disconnect between the oscillations of the optical and mechanical degrees of freedom. Indeed, both mechanical variables $q_y$ and $q_z$ exhibit nearly harmonic oscillations close to the bare mechanical frequency $\Omega$, with $q_y$ remaining somewhat more harmonic than $q_z$, whose dynamics is more strongly affected by coupling to $J_z$.

At the same time, the pseudospin components $J_x$ and $J_z$ oscillate at substantially higher frequencies, interrupted by sharp modulation events occurring approximately every half mechanical period. The component $J_y$ exhibits similar dynamics but with a significantly smaller amplitude. Thus, in contrast to the small-$q_0$ regime, the mechanical oscillation now defines the slow time scale of the dynamics.

The behavior of the optical variables can be understood as precession of the vector $\mathbf{J}$ with frequency
\begin{equation}\label{eq:precession}
\Lambda=\sqrt{\left(\Delta-\sqrt{2}g_1^zq_z\right)^2+4g_y^2q_y^2}
\end{equation}
around the effective pseudo-magnetic field $\mathbf{\Lambda}$, Eq.~\ref{eq:magnetic_field}. For sufficiently large $q_0$, such that $g_y q_0 \gg \Delta \sim 1$, the precession frequency becomes much larger than the mechanical frequency. In this regime the mechanical displacements $q_y$ and $q_z$ act as slowly varying adiabatic parameters that modulate the precession frequency.

The comparatively small amplitude of the $J_y$ oscillations reflects the fact that the effective field $\mathbf{\Lambda}$ is oriented predominantly along the $y$-direction, so that the corresponding precession primarily involves oscillations of the $J_x$ and $J_z$ components. 
The return of the mechanical degrees of freedom to oscillations close to their bare frequency can be understood as a consequence of rapid pseudospin precession. Since the components of $\mathbf{J}$ oscillate on a much faster time scale than the mechanical motion, their effect on the mechanical dynamics becomes largely self-averaging. As a result, the optical contribution to the motion in the $y$-direction reduces primarily to a small shift of the oscillation baseline, while in the $z$-direction it produces only weak corrections to both the oscillation baseline and the effective mechanical frequency (see details in Methods section~\ref{sub:unstable}).

While this picture is helpful for a qualitative understanding of this regime, for the chosen parameter values $q_0=3\times10^5$ and $g_y=3\times10^{-5}$ the separation between the mechanical and optical frequencies is rather marginal, as we observe only about ten optical oscillations during a single mechanical period. It is therefore remarkable that, even though the adiabatic argument applies only qualitatively, the self-averaging of the optical oscillations remains highly effective.

The abrupt changes in the otherwise well-defined oscillations of the optical variables occur twice during each mechanical period  and coincide with the mechanical variables passing through zero. This behavior reveals a close analogy between the present model and the Landau–Zener problem~\cite{Landau1932,Zener1932,Stueckelberg1932,Majorana1932}, which describes nonadiabatic transitions between states of a two-level quantum system.

Indeed, the equations of motion for the expectation values of the pseudospin  operators can be generated by a Hamiltonian of the form
\begin{equation}\label{eq:Landau-Zener}
\hat{H}= -\frac{1}{2}\hbar\, \mathbf{\Lambda}\cdot\hat{\mathbf{\sigma}},
\end{equation}
where $\hat{\mathbf{\sigma}}$ are the Pauli matrices (see Methods section~\ref{sub:unstable}). In the basis of the eigenvectors of $\sigma_y$, this Hamiltonian becomes analogous to the standard Landau--Zener Hamiltonian~\cite{Landau1932,Zener1932,Stueckelberg1932,Majorana1932}. In this representation, the uncoupled ``energy levels'' of the effective two-level system are given by $\pm\sqrt{2}g_y q_y$, while the coupling between them is determined by $g_1^z q_z-\Delta$. In the vicinity of the zeros of $q_y$, one may approximate $q_y\propto t$, which transforms Hamiltonian~\eqref{eq:Landau-Zener} into the standard Landau–Zener form. The interruptions of the adiabatic dynamics observed in Fig.~\ref{fig:unstable-large-q} can therefore be interpreted as nonadiabatic transitions arising from the breakdown of the adiabatic condition when the separation between the two effective ``energy levels'' becomes too small.

\begin{figure}
\centering
\begin{subfigure}{\textwidth}
         \centering \includegraphics[width=\textwidth]{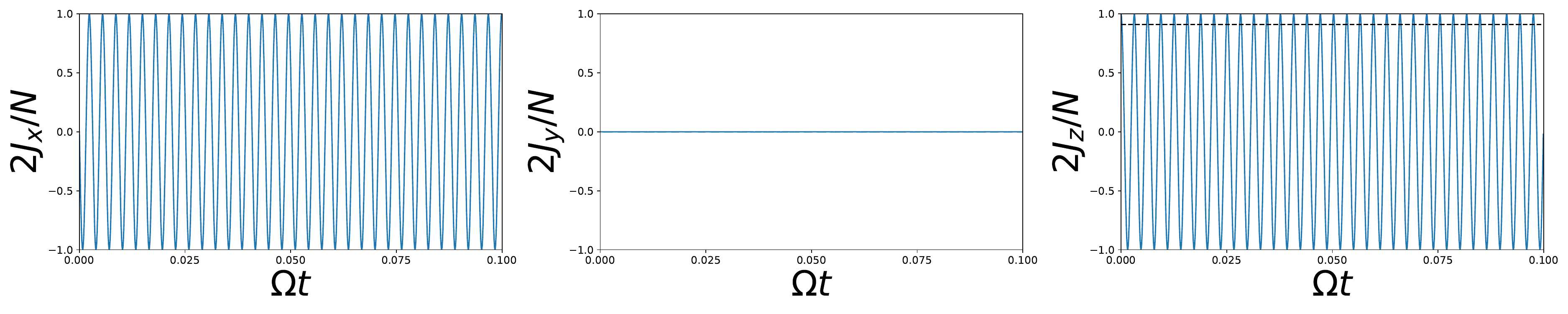}
\end{subfigure}
\vfill
\begin{subfigure}{\textwidth}
         \centering \includegraphics[width=\textwidth]{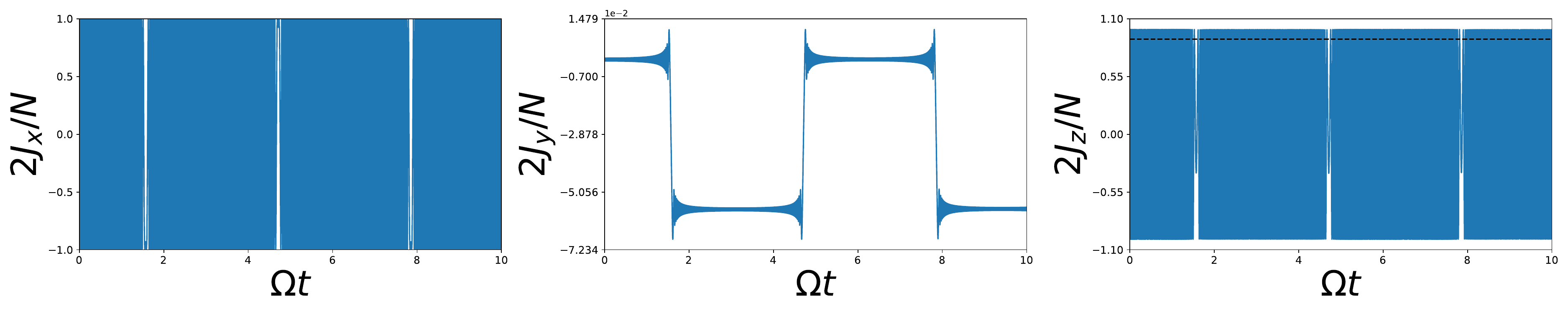}
\end{subfigure}
\vfill
\begin{subfigure}{\textwidth}
         \centering \includegraphics[width=\textwidth]{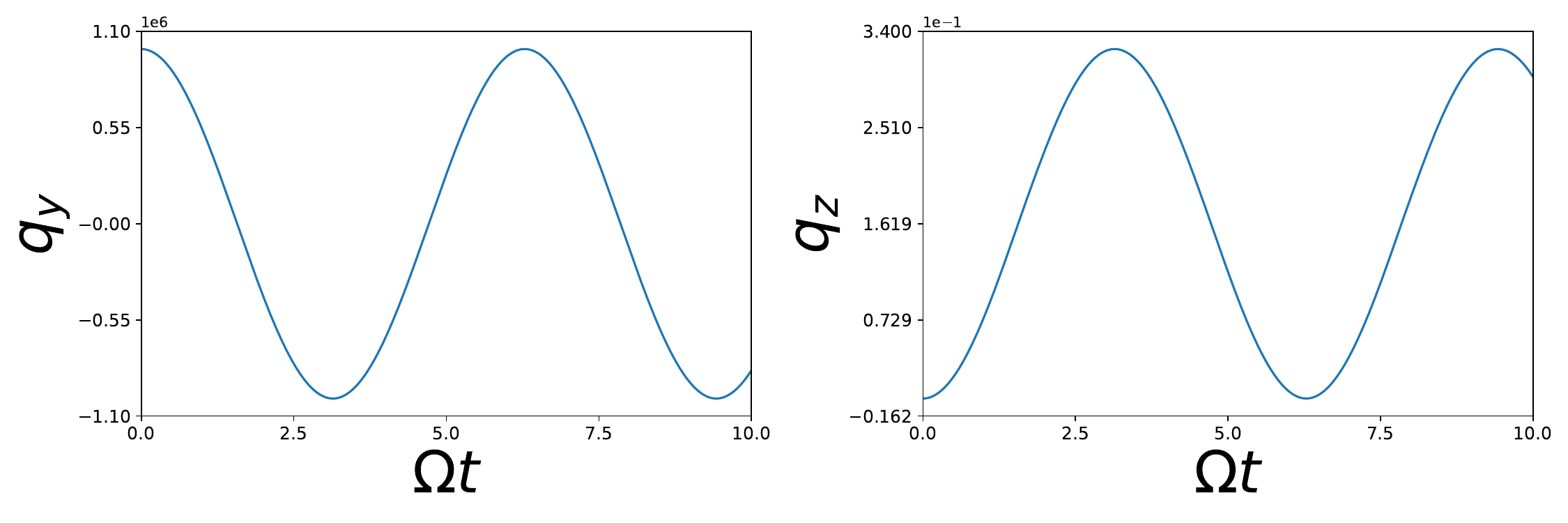}
\end{subfigure}
\caption{Evolution of optical ($J_x,J_y,J_z$) and mechanical ($q_y,q_z$) variables in the unstable regime corresponding to strongly non-linear case with $q_0=10^6$ and $N/N_c=1.1$ and $g_y=10^{-3}$. }
\label{fig:unstable-large-q-g}
\end{figure}

The separation of time scales between the optical and mechanical degrees of freedom becomes more pronounced if the parameters are increased to $g_y=10^{-3}$ and $q_0=10^6$, thereby increasing the optical precession frequency relative to the mechanical frequency. While such parameters may not be directly accessible in the WGM realization, it is nevertheless instructive to illustrate this regime (see Fig.~\ref{fig:unstable-large-q-g}). The first row in this figure shows oscillations of the optical variables over a time interval much shorter than the mechanical period. The components $J_{x,z}$ undergo multiple fast oscillations, while $J_y$ remains nearly constant. The near constancy of $J_y$ indicates that the vector $\mathbf{\Lambda}$ is now oriented predominantly along the $y$-axis. 

The second row shows the same variables over a time interval covering several mechanical periods, where one can clearly see the interruptions of the adiabatic behavior. They are especially pronounced in the dynamics of $J_y$, which abruptly changes sign, corresponding to a reversal of its orientation relative to the instantaneous rotation axis. The last row of the figure shows the mechanical variables, which exhibit nearly harmonic oscillations at the frequency $\Omega$.

\begin{figure}
\centering
\vspace{0\linewidth}
\includegraphics[width=0.5\linewidth]{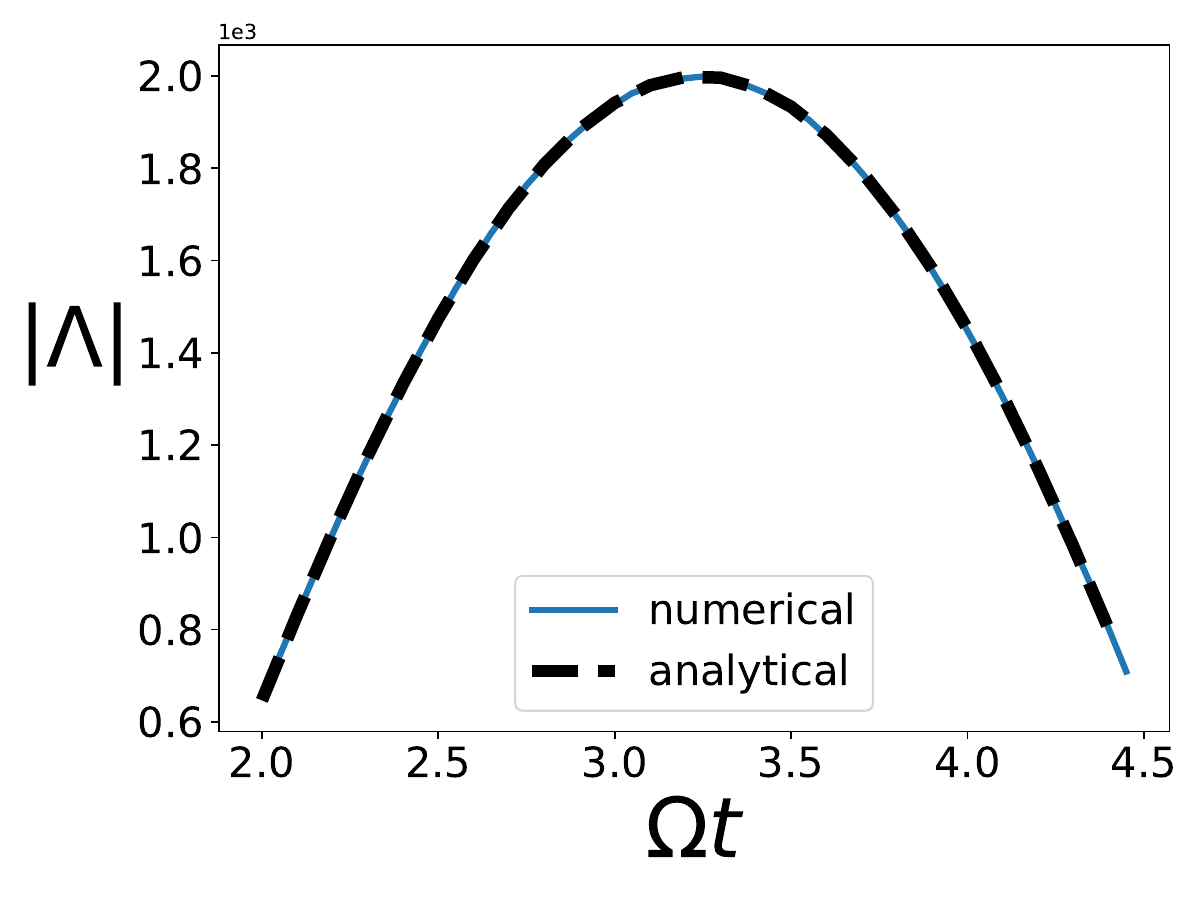}
\caption{Time variation of the optical frequency obtained numerically (for $N/N_c=1.1$, $g_y=10^{-3}$ and $q_0=10^6$, binned in $0.1$ s intervals) and compared with the adiabatic prediction  $\Lambda$ described by Eq.~\ref{eq:precession}.}
\label{fig:mod_freq_unstable}
\end{figure}

To assess the accuracy of the adiabatic approximation underlying this description, we divided the optical energy evolution into time intervals much shorter than the mechanical period and performed a local Fourier analysis within each interval. The resulting dominant instantaneous frequency agrees well with the adiabatic prediction given by Eq.~\ref{eq:precession} (see Fig.~\ref{fig:mod_freq_unstable}).

It is important to emphasize that the bifurcation, and therefore the associated instability, remains robust in the presence of weak optical dissipation. As long as the positive imaginary part of the unstable eigenfrequency exceeds the optical decay rate, the dynamical effects discussed in this section persist. Since the instability increment increases with the initial number of photons injected into the $m=0$ mode, the instability can, in principle, be tuned to overcome moderate optical losses.

\subsection{Summary and Outlook}
The multimode optomechanical systems with transverse intermode coupling considered in this work complement the established paradigm of cavity optomechanics by providing access to dynamical regimes that are inaccessible in systems dominated by conventional longitudinal dispersive or dissipative coupling. In the present paper, such transverse coupling emerges from a mechanically induced symmetry-breaking mechanism. However, many of the dynamical phenomena discussed here are expected to be generic for a broader class of systems with transverse optomechanical interactions. 

The proposed coupling mechanism may be particularly relevant in systems with soft mechanical modes, such as droplet resonators, where mechanical motion is represented by capillary oscillations~\cite{Dahan2016droplet,Bahl2012NatPhys}. In such systems, the low mechanical frequencies make realization of the resolved-sideband regime difficult, limiting access to many standard dynamical effects of longitudinal optomechanics. Under these conditions, dynamical effects associated with transverse optomechanical coupling may become comparatively more important.

By selectively exciting different optical modes one can generate distinct nonlinear dynamical regimes. In this work we analyzed one such configuration, which exhibits two qualitatively different regimes separated by a bifurcation. Different dynamical regimes may enable qualitatively different functionalities, ranging from alternative cooling mechanisms beyond the conventional sideband-resolved paradigm to mechanical control of optical frequency conversion processes.

In this work we focused primarily on Hamiltonian (dissipationless) dynamics. While optical decay suppresses dynamical effects developing on time scales longer than the cavity lifetime, the conservative bifurcation condition is expected to remain relevant as long as the instability increment exceeds the optical decay rate. Numerical simulations presented in the Supplementary Materials demonstrate that even in the bad-cavity regime the high-frequency optical oscillations and associated broad sideband structure survive, while slower low-frequency dynamics becomes progressively suppressed by dissipation. Experimental access to these short-time coherent effects may be further extended using stroboscopic ring-down excitation protocols, conceptually similar to dynamical-decoupling techniques developed in quantum control~\cite{Viola1998}.

To demonstrate robustness of the short-time nonlinear effects we simulated nonlinear behavior of our system in the stable regime demonstrating that even under bad cavity conditions, the high frequency sidebands generated by mechanical modulation of the intermode coupling do survive. 

The numerical examples presented in this paper were based on a specific experimental realization—a spherical droplet resonator interacting with a functionalized nanobeam cantilever. Our primary goal was not to provide precise experimental predictions, but rather to demonstrate that the parameter regime required for the onset of the predicted nonlinear dynamics is not fundamentally beyond the reach of existing experimental platforms. In particular, we sought to keep the critical intracavity power associated with the bifurcation transition within the sub-milliwatt range. Since the present work focuses primarily on conservative dynamics, these estimates should be viewed as illustrative order-of-magnitude evaluations rather than quantitatively optimized experimental conditions. This requirement nevertheless imposed significant constraints on the choice of working parameters such as $\Delta$, $g_{y,z}$, and the initial displacement $q_0$. If the goal is instead to operate in the stable regime, these restrictions can be relaxed and the ratio $g_y/\Omega$ can be significantly increased, potentially extending the optical span of the generated sidebands (Fig.~\ref{fig:stable-evolution-high}) even further.

In summary, this work represents a first step toward an alternative direction in cavity optomechanics based on symmetry-breaking transverse coupling, opening broad possibilities for further theoretical and experimental exploration. 

\section{Method}\label{sec:method}

\subsection{ Hamiltonian and Schwinger operators}\label{subsec:equation_motion}

The symmetry-breaking interaction term in the Hamiltonian presented in Eq.~\ref{eq:Hamiltonian-dark-mode} possesses an intrinsic $SU(2)$ structure, which becomes explicit upon introducing the Schwinger pseudospin operators defined in Eq.~\ref{eq:Schwinger-operatorsJx}--\ref{eq:Schwinger-operatorsJz}.

The Hamiltonian now can be rewritten as
\begin{equation}
    \begin{split}
        \hat{H}=&-\hbar\Delta\left(\frac{1}{2}\hat{N}_{b}+\hat{N}_d-\hat{J}_{z}\right)+\hbar\Omega\left(\hat{b}_{y}^{\dagger}\hat{b}_{y}+\hat{b}_{z}^{\dagger}\hat{b}_{z}\right)\\+
&\hbar\left(\hat{b}_{z}^{\dagger}+\hat{b}_{z}\right)\left[\frac{1}{2}N_{b}\left(g_{1}^{z}+g_{0}^{z}\right)+g_{1}^{z}\hat{N_d}-\hat{J}_{z}\left(g_{1}^{z}-g_{0}^{z}\right)\right]\\+
&\sqrt{2}\hbar g_y\left(\hat{b}_{y}^{\dagger}+\hat{b}_{y}\right)\hat{J}_{y}.
    \end{split}
\label{eq:Hamiltonian Schwinger}\end{equation} 
Here $\hat{N}_b=\hat{\alpha}^\dagger\hat{\alpha}+\hat{a}_0^\dagger\hat{a}_0$ is a conserving number of photons in the bright (interacting) optical modes, and $\hat{N}_d=\hat{\beta}^\dagger\hat{\beta}$ is a conserving number of photons in the dark mode. The second line in Eq.~\ref{eq:Hamiltonian Schwinger} describes the traditional dispersive coupling and contains only $J_z$. In the language of cavity QED, this interaction corresponds to longitudinal coupling. The last line describes the symmetry-breaking interaction proportional to $J_y$, corresponding to transverse coupling in the cavity-QED terminology.  The corresponding Heisenberg equations for dimensionless mechanical coordinates and momentum defined in Eq.\ref{eq:q_p_dimensionless}
can be written in the following form
\begin{eqnarray}
    \frac{d\hat{q}_{z}}{dt}&	=	& \hat{p}_{z};\label{eq:qz}\\
\frac{d\hat{p}_{z}}{dt}&	=&	- \hat{q}_{z}-\sqrt{2}\left[\frac{N_{b}}{2}\left(g_{1}^{z}+g_{0}^{z}\right)+g_{1}^{z}N_{d}-\hat{J}_{z}\left(g_{1}^{z}-g_{0}^{z}\right)\right]\label{eq:pz}\\ 
\frac{d\hat{q}_{y}}{dt}	&=	& \hat{p}_{y}\label{eq:qy}\\
\frac{d\hat{p}_{y}}{dt}&	=&	- \hat{q}_{y}-2g_{y}\hat{J}_{y},\label{eq:py}
\end{eqnarray}
where all quantities with dimensions of frequency or time are now expressed in units of $\Omega$ or $1/\Omega$ respectively, but retain the same notation for simplicity. 

The equations for  the pseudospin operators $\hat{J}_i$ representing optical variables have the following form 
\begin{eqnarray}
\frac{d\hat{J}_{x}}{dt}&	=&	-\left[\Delta-\sqrt{2}\left(g_{1}^{z}-g_{0}^{z}\right)q_{z}\right]\hat{J}_{y}+2g_{y}q_{y}\hat{J}_{z};\label{eq:Jx}\\
\frac{d\hat{J}_{y}}{dt}&	=&	\left[\Delta-\sqrt{2}\left(g_{1}^{z}-g_{0}^{z}\right)q_{z}\right]\hat{J}_{x};\label{eq:Jy}\\
\frac{d\hat{J}_{z}}{dt}&	=&	-2g_{y}q_{y}\hat{J}_{x};\label{eq:Jz}
\end{eqnarray}
and can be written in the form of the effective Bloch equations \begin{equation}\label{eq:Bloch}
    \frac{d\hat{\mathbf{J}}}{dt}=\hat{\mathbf{J}}\times \hat{\mathbf{\Lambda}}
\end{equation}
with $\hat{\mathbf{\Lambda}}$ defined as
\begin{equation}
   \hat{\Lambda}_x=0, \hspace{2pt}\hat{\Lambda}_y=2g_yq_y,\hspace{2pt} \hat{\Lambda}_z=-\left[\Delta-\sqrt{2}\left(g_1^z-g_0^z\right)\hat{q}_z\right]
\end{equation}

In the classical limit, all operators in Eqs.~\ref{eq:qy}--\ref{eq:Bloch} are replaced by their expectation values, after which the $\hat{ }$ decoration is dropped.

The fixed point of these equations is given by
\begin{equation}
\begin{split}
   & p_z^{eq}=p_y^{eq}=q_y^{eq}=J_y^{eq}=J_x^{eq}=0;\\
   & q_z^{eq}=-\frac{1}{\sqrt{2}}\left[N_b\left(g_1^z+g_0^z\right)+2g_1^zN_d-2\left(g_1^z-g_0^z\right)J^{eq}_z\right].
    \end{split}
\end{equation}

The equilibrium value of $J_z$ is determined by the total number of photons in the bright optical subsystem, $N_b$. Indeed, the Schwinger operators satisfy the standard angular-momentum algebra, implying that the eigenvalues of $\hat{J}_z$ range from $-N_b/2$ to $N_b/2$, while the eigenvalues of $\hat{\mathbf{J}}^2$ are given by $N_b/2(N_b/2+1)$. Since the fixed point satisfies $J_x=J_y=0$, the pseudospin vector is fully polarized along the $z$-axis and therefore $J_z^{eq}=\pm N_b/2$ in the classical limit. The sign of $J_z^{eq}$ determines the stability of the optical subsystem. To see this, we note that Eq.~\ref{eq:Bloch} can be reproduced as the Heisenberg or Ehrenfest equations generated by the effective Hamiltonian
\begin{equation}\label{eq:BlochHamilt}
    \hat{H}=-\hat{\mathbf{\Lambda}}\cdot \hat{\mathbf{J}}.
\end{equation}
The state with $J_z=N_b/2$ corresponds to the maximum of this Hamiltonian with respect to the optical degrees of freedom and is therefore optically unstable, while $J_z=-N_b/2$ corresponds to the stable optical equilibrium. In what follows we consider the unstable equilibrium $J_z^{eq}=N_b/2$, for which the equilibrium displacement takes the form
\begin{equation}\label{eq:qz-eq}
q_z^{eq}=-\frac{1}{\sqrt{2}}\left[N_bg_0^z+2g_1^zN_d\right].
\end{equation}

Linearization of these equations around this fixed point results in two separate systems: the longitudinal mechanical variables $q_z$ and $p_z$ decouple into separate equations 
\begin{eqnarray}
    \frac{dq_{z}}{dt}&	=	& p_{z};\\
\frac{dp_{z}}{dt}&	=&	- q_{z}-\frac{1}{\sqrt{2}}\left[N_{b}\left(g_{1}^{z}+g_{0}^{z}\right)+2g_{1}^{z}N_{d}-2J_{z}\left(g_{1}^{z}-g_{0}^{z}\right)\right]
\end{eqnarray} 
while transverse and pseudospin variables form a separate system of equations
\begin{equation}\label{eq:linear}
    \begin{split}
        \frac{d{q}_{y}}{dt}	&=	{p}_{y}\\
\frac{d{p}_{y}}{dt}	&=	-{q}_{y}-2g_{y}J_{y}\\
\frac{dJ_{x}}{dt}&	=	-\Delta_z J_{y}+2g_{y}J_{z}^{eq}q_y\\
\frac{dJ_{y}}{dt}&	=	\Delta_z J_{x}
    \end{split}
\end{equation}
where $\Delta_z$ is defined as 
$$
\Delta_z=\Delta-\sqrt{2}(g_1^z-g_0^z)q_z^{eq},
$$ 
and $J_z$ is replaced with its equilibrium value $J_z=J_z^{eq}$. Under the assumption introduced in Section~\ref{sec:results} ($g_0^z=0$, $N_d=0$) 
$$
q_z^{eq}=0
$$ and $\Delta_z=\Delta$. The system of equation given by Eq.~\ref{eq:linear} generates coupled optomechanical oscillations with eigenvalues presented in Eq.~\ref{eq:eigenfreq}. 

To simulate the nonlinear dynamics of our model we chose simple initial conditions
\begin{equation}
    q_z(0)=p_z(0)=p_y(0)=J_y(0)=J_x(0)=0;\hspace{2pt} q_y(0)=q_0;\hspace{2pt}J_z(0)=N_b/2=N/2
\end{equation}
It shall be noted that an alternative choice with $q_y=0$ and $q_z\ne 0$ wouldn't generate any nontrivial dynamics since the transverse degrees of freedom in this case would not be excited. 

Transitioning back to optical amplitudes $a_0$ and $\alpha$ we reproduce initial conditions presented in Eq.~\ref{eq:initial_condition}. 

\subsubsection{ Stable regime}\label{sub:stable}

To quantify the dynamics in the stable regime, we compute the normalized optical and mechanical energies (measured in units of $\hbar\Omega$):
\begin{eqnarray}
 E_{opt}&=&-\Delta\left(\frac{1}{2}N_b-J_z\right)\equiv-\Delta\alpha^\dagger\alpha,\label{eq:ene-opt}\\
 E_{mech}&=&(q_y^2+p_y^2)/2+(q_z^2+p_z^2)/2\label{eq:ene-mech}.
\end{eqnarray}

For the initial conditions given in Eq.~\ref{eq:initial_condition}, the linear dynamics consists of superpositions of normal modes with eigenfrequencies $\pm\lambda$ and $\pm\nu$. Since the energies are quadratic in the dynamical variables, their spectra contain oscillations at frequencies $2\lambda$, $2\nu$, and combinations $\lambda\pm\nu$. In the regime $\Delta\sim\Omega$, the frequencies $\lambda$ and $\nu$ are close to each other, so that $\lambda-\nu\ll\lambda$. As a consequence, the energy dynamics is dominated by slow oscillations at the difference frequency $\lambda-\nu$.

In the regime $N_b\ll N_c$ and $\Delta\gg\Omega$, illustrated in Fig.~\ref{fig:stable-evolution-low_old}, one finds $\lambda\gg\nu$. The modes oscillating at frequencies $\pm\nu$ remain only weakly affected by nonlinear effects, while the dynamics associated with $\pm\lambda$ dominates the optical and mechanical energies. Consequently, $E_{opt}(t)$ and $E_{mech}(t)$ exhibit approximately harmonic oscillations with slowly varying amplitudes.

This modulation can be derived by transforming Eqs.~\ref{eq:qz}--\ref{eq:Jz} into the basis of normal modes (see Supplementary Materials for details). The mode associated with $\lambda$ takes the form
\[
\psi_1=-\frac{g_y q_0}{2\lambda}\big[r_2(t)\cos(\lambda t)+r_1(t)\sin(\lambda t)\big],
\]
with $r_1(0)=1$ and $r_2(0)=0$. Averaging over the fast oscillations at frequency $\lambda$ yields
\begin{equation}
     \frac{d^2r_{1,2}}{dt^2}+\frac{g_y^2q_0^2}{8}r_{1,2}=0,
\end{equation}
which leads to the modulation frequency given in Eq.~\ref{eq:modulation_freq}.

\subsubsection {Unstable regime}\label{sub:unstable}

The main feature of the strongly nonlinear unstable regime is the effective decoupling between nearly harmonic mechanical oscillations occurring at the bare mechanical frequency and substantially faster optical oscillations. This separation of time scales allows the dynamics to be understood qualitatively within an adiabatic approximation. In this picture, the optical dynamics corresponds to precession of the pseudospin vector $\mathbf{J}$ governed by the Bloch equation, Eq.~\ref{eq:Bloch}, with precession frequency given by the magnitude of the effective pseudomagnetic field $\mathbf{\Lambda}$ defined in Eq.~\ref{eq:precession}. For large enough initial displacements $q_y$, and $q_z\ll q_y$ we can reach the regime when 
 \begin{equation}\label{eq:adiabatic}
     g_y q_y \gg \left|\Delta-\sqrt{2}g_1^z q_z\right| \gg 1,
 \end{equation}
 and the optical modulation frequency is therefore predominantly determined by the large mechanical displacement $q_y$. 
 
 An adiabatic solution of Eq.~\ref{eq:precession} with ``frozen'' mechanics is obtained using a rotation matrix $\hat{R}$ corresponding to rotation around the unit vector $\mathbf{n}=\{0,n_y,n_z\}$ (parallel to $\mathbf{\Lambda}$) through an angle $\Lambda t$, i.e., $\mathbf{J}=\hat{R}\mathbf{J}_0$ (the explicit form of $\hat{R}$ is given in the Supplementary Materials). To understand why the mechanical oscillations remain close to their bare frequency, we decompose the rotating pseudospin vector into components parallel and perpendicular to the rotation axis: $J_n=\mathbf{J}\!\cdot\!\mathbf{n}$ and $\mathbf{J}_\perp$. The transverse component $\mathbf{J}_\perp$ averages to zero over fast rotation, while the longitudinal component $J_n\equiv u$ is an adiabatic invariant. Hence, the cycle-averaged optical pseudospin  vector is $\overline{\mathbf{J}}=u\,\mathbf{n}$, which gives
\begin{equation}\label{eq:time-averaged}
        \overline{J}_y=-\frac{2g_y q_y}{\Lambda}\,u;\hspace{4 pt}
        \overline{J}_z=-\frac{\Delta-\sqrt{2}g_1^z q_z}{\Lambda}\,u.
\end{equation}
Under the condition of Eq.~\ref{eq:adiabatic}, we have $\overline{J}_y\approx u$. Substituting Eq.~\ref{eq:time-averaged} into Eqs.~\ref{eq:py} and \ref{eq:pz} shows that $\overline{J}_y$ primarily produces a quasistatic shift of the equilibrium position of $q_y$ without significantly affecting its oscillation frequency, consistent with the numerical simulations. At the same time, $\overline{J}_z$ generates only small corrections to the effective frequency and equilibrium position of $q_z$, of order $u(g_1^z)^2N_b/\Lambda$ and $u g_1^z N_b \Delta/\Lambda$, respectively. Both effects become negligible in the adiabatic limit. Both effects are negligible in the adiabatic limit. 

To elucidate the origin of the nonadiabatic jumps occurring twice during each mechanical period, we note that although the Hamiltonian in Eq.~\ref{eq:BlochHamilt} formally describes a pseudospin of magnitude $N_b/2$, the same equations of motion for the expectation values of the pseudospin components can be generated by the effective spin-$\frac{1}{2}$ Hamiltonian given in Eq.~\ref{eq:Landau-Zener}. This reduction remains valid as long as the pseudospin variables are treated as expectation values rather than full quantum operators. In the basis of the eigenvectors of the y-component of the pseudospin, this effective Hamiltonian takes the matrix form 
\begin{equation}
    \hat{H}=
    \begin{bmatrix}2g_yq_y & -\left[\Delta -\sqrt{2}g_1^zq_z\right]\\
-\left[\Delta -\sqrt{2}g_1^zq_z\right] & -2g_yq_y
\end{bmatrix}
\end{equation}
which is identical in structure to  the standard Landau-Zener Hamiltonian. The non-adiabatic transitions observed in our simulations can therefore be interpreted as Landau–Zener–type events arising from the breakdown of adiabaticity when the mechanical displacement $q_y(t)$ passes through zero. In the vicinity of the zeros of $q_y$, one may approximate $q_y\propto t$, so that the Hamiltonian in Eq.~\ref{eq:Landau-Zener} locally reduces to the standard Landau--Zener form, with uncoupled instantaneous energies $\pm \sqrt{2}\,g_y q_y$ and coupling $g_1^z q_z-\Delta$. 

\subsection{Potential experimental realization of the model}\label{subsec:experiment}

The main parameters determining the feasibility of experimental observation of the effects discussed in this paper are the dispersive and transverse coupling constants $g_{1,0}^z$ and $g_y$, together with the critical photon number $N_c$. Equation~\ref{eq:Nc} indicates considerable flexibility in designing systems spanning a broad range of values of $N_c$. In particular, varying the detuning between the optical mode splitting $\Delta$ and the mechanical frequency $\Omega$ allows the bifurcation threshold to be tuned over a wide range. At exact resonance, $\Delta=\Omega$, the critical photon number vanishes. 

Here we outline a potential implementation of our model based on a spherical  whispering-gallery-mode (WGM) resonator interacting with a nanobeam cantilever and provide experimentally feasible values for the relevant quantities.  

The high degree of spherical symmetry required by the model can be realized  in liquid droplets levitated in air or water using optical or acoustic traps (see, for instance \cite{KherAlden2020}). One possible realization is a PDMS droplet. In the large-$L$ limit, the azimuthal mode number can be estimated as
\begin{equation}
L \simeq \frac{2\pi n R}{\lambda_0},
\end{equation}
where $\lambda_0$ is the vacuum wavelength at the frequency of the WGM resonance and  $n$ is the refractive index of the droplet. 

The mechanical oscillator can be realized as a nanobeam cantilever functionalized by a small ($\sim 30~\mathrm{nm}$) dielectric particle. The relevant mechanical parameters are the effective mass $m_{\mathrm{eff}}$ and eigenfrequency $\Omega$. A small size of the particle, which is much smaller than relevant optical wavelength allows to treat it as a polarizable dipole\cite{Deych2009}. A tip of such a cantilever positioned at a distance $d$ from the surface of the resonator induces a separation of frequencies between previously degenerate modes $\Delta$, which can be estimated as  
\begin{equation}\label{eq:freq_splitting}
\Delta(d) \simeq \frac{\omega_c}{2}
\frac{\alpha_{\rm eff}}{V_{\rm eff}}
\, e^{-2\kappa d},
\qquad
\omega_c = \frac{2\pi c}{\lambda},
\end{equation}
where $\alpha_{\rm eff}$ is the effective polarizability of the tip and $V_{\rm eff}$ is the optical mode volume. The evanescent decay constant is
\begin{equation}
\kappa \simeq k_0
\sqrt{n^2 - n_{\rm out}^2},
\end{equation}
with $n_{\rm out}$ the refractive index of the surrounding medium.

The symmetry-breaking optomechanical coupling parameters for this model was presented in Eq.~\ref{eq:transv_coupling}, while the dispersive coupling parameter $g_1^z$ can be estimated using Eq.~\ref{eq:disp_coupling} and Eq.~\ref{eq:freq_splitting} as 
\begin{equation}
g_1^z \simeq -2\kappa\,\Delta\,x_{zpf}.
\end{equation}
Furthermore we set $g_0^z=0$, which is consistent with TE polarized WGM modes interacting with a dipole\cite{Deych2009}. In what follows we will simplify notation by replacing $g_1^z$ with $g_z$.  Dynamics of the system  depends sensitively on the ratio of the two coupling parameters, which in the WGM setup is  
\begin{equation}
\frac{g_z}{g_y}
\simeq \frac{2\kappa R}{L}
\simeq 2\,\frac{\sqrt{n^2 - n_{\rm out}^2}}{n}.
\end{equation}
This ratio can therefore be either smaller or larger than unity, depending on the refractive-index contrast between the resonator and the surrounding medium. 

The critical intracavity photon number is given by Eq.~\ref{eq:Nc}. Substituting the expressions for $g_y$ and $x_{zpf}$ into Eq.~\ref{eq:Nc} yields
\begin{equation}
N_c \simeq
\frac{m_{\rm eff} \lambda^2 \Omega}
     {16\pi^2 n^2 \hbar}
\frac{(\Delta^2 - 1)^2}{\Delta^3},
\end{equation}

As a representative system we consider a droplet  of radius $R \sim 20~\mu{\rm m}$ at wavelength
$\lambda \sim 780~{\rm nm}$ with refractive index $n \simeq 1.4$.
The order of the corresponding WGM can be estimated as $L \sim 200$ and the spatial decay parameter of the evanescent field
$\kappa \sim 10^7~{\rm m}^{-1}$.

For a cantilever with a tip with an effective polarizable size of a few tens of nanometers
and a nanoscale tip--droplet separation, the above estimate yields
\begin{equation}
\Delta \sim 10^8~{\rm s}^{-1}.
\end{equation}

For an ultralight cantilever we can realistically take 
\begin{equation}
m_{\rm eff} \sim 10^{-19}~{\rm kg},
\qquad
\Omega \sim 2\times 10^8~{\rm s}^{-1}.
\end{equation}
In this case the zero-point amplitude becomes
\begin{equation}
x_{zpf} \simeq 2.3\times 10^{-12}~{\rm m}.
\end{equation}
and the corresponding coupling parameters become 
\begin{equation}
g_y \simeq -6\times 10^3~{\rm s}^{-1},
\qquad
g_z \sim -8.4\times 10^3~{\rm s}^{-1}.
\end{equation}
The dimensionless version of these parameters used in our numerical simulations is then $-3\times 10^{-5}$ and $-4.2\times 10^{-5}$ respectively.

To reach the nonlinear regime the value of the initial dimensionless displacement $q_y(0)$ must be chosen of the order $10^4$. Taking into account that the length scale in our model is established by $x_{zpf}$ such values correspond to displacement of the order of $0.01 \mu m$, which remain within the linear displacement regime of a typical cantilever.   
Finally, the critical photon number is estimated as
\begin{equation}
N_c \sim 10^5\text{--}10^6.
\end{equation}
which corresponds to the intracavity optical power of the order of $\mu W$. It should be emphasized that these numbers are given for illustrative purposes only to demonstrate that the  phenomena presented in this paper can be in principle observed under realistic experimental conditions. The present estimates are not intended to represent optimized device parameters and could potentially be improved in alternative photonic geometries.

\section*{Acknowledgments}
Authors acknowledge the help of Dr. Mishkatul Bhattacharya and Kewen Xiao, who participated at the earlier stages of this work. The authors also thank the anonymous referee for critical comments and suggestions that helped improve the clarity and presentation of this work. This work was funded by NSF award 2102249. 
\bibliography{bibfile} 
\bibliographystyle{apsrev4-1}
\end{document}


\title{Supplementary materials to "Transverse Optomechanical Interaction Mediated by Mechanically Induced Symmetry Breaking: Hamiltonian Dynamics"}

\author{Satyam S. Jha}
\affiliation{Department of Physics, Queens College of CUNY, Flushing, NY 11367}
\affiliation{Graduate Center of CUNY, 365 5th Ave, New York, NY 10016}

\author{Lev Deych}
\thanks{Author to whom correspondence should be addressed: lev.deych@qc.cuny.edu}
\affiliation{Department of Physics, Queens College of CUNY, Flushing, NY 11367}
\affiliation{Graduate Center of CUNY, 365 5th Ave, New York, NY 10016}

\maketitle

\section{Derivation of Hamiltonian with transverse interaction for spherical WGM resonator interacting with a polarizable dipole}
The frequencies of the affected modes depend only on the radial coordinate of the oscillator, which in the chosen coordinate system is also its Cartesian $z$-coordinate. 
However, to account for the possibility of particle's motion, we need to rewrite this Hamiltonian in a stationary coordinate system not attached to the position of the particle. We chose a polar axis in the direction in which angular coordinates of the particle are some arbitrary $\theta,\varphi$ so that the new Z axis is defined by the unit vector 
\begin{equation}\label{eq4}
\boldsymbol{e}_{Z}=\boldsymbol{e}_{z}\cos\theta+\boldsymbol{e}_{x}\sin\theta\cos\varphi+\boldsymbol{e}_{y}\sin\theta\sin\varphi.
\end{equation}
The choice of two other axes is arbitrary, so we define the new Y axis as
\begin{equation}\label{eq5}
   \boldsymbol{e}_{Y}=-\boldsymbol{e}_{x}\sin\theta\sin\varphi+\boldsymbol{e}_{y}\sin\theta\cos\varphi, 
\end{equation}
and the direction of the X axis is given by $\boldsymbol{e}_{X}=\boldsymbol{e}_{Y}\times\boldsymbol{e}_{Z}$. The standard Euler's angles $\alpha,\beta$, and $\gamma$, as defined, for instance in \cite{Mishchenko2002} and describing the rotation between the “old” and the “new” coordinate systems can be found as 
\begin{equation}\label{eq:Euler_angles}
\beta=\theta,\:\alpha=\varphi+\pi/2,\:\gamma=\pi/2
\end{equation}
Using transformation properties of the spherical harmonics we can introduce creation and annihilation operators defined in the rotated stationary coordinate system and the corresponding Hamiltonian:
\begin{equation}\label{eq:Hamiltonian-global}
\hat{H}=\sum_{m_{1},m_{2}}J_{m_{1},m_{2}}\left(r,\theta\right)\hat{a}_{L,m_{1}}^{\dagger}\hat{a}_{L,m_{2}}
\end{equation}
where 
\begin{equation}\label{eq9}
J_{m_{1},m_{2}}\left(r,\theta\right)=\left(-1\right)^{m_{2}}\left(i\right)^{m_{2}+m_{1}}\\
\times \sum_{m}\omega_{L,m}\left(r\right)d_{mm_{1}}^{L}\left(\theta\right)d_{mm_{2}}^{L}\left(\theta\right)\\
\end{equation}
and $d_{mm_{1}}^{L}\left(\theta\right)$ is the Wigner small $d$-matrix. This Hamiltonian is now non-diagonal in the azimuthal numbers, which reflects the physical fact that the motion of the particle results in such reconfiguration of the field that it can't be described by a single spherical harmonics in a stationary coordinate system. This Hamiltonian describes two different types of optomechanical interaction: in addition to the standard interaction due to the position dependence of the resonator's frequency, there is an additional interaction due to the coupling between different degenerate modes of the resonator.

The Wigner small $d$-matrix has the following form as presented in Ref.\cite{Mishchenko2002} 
\begin{equation}\label{eq:d-matrix}
\begin{split}
    d_{mm_{1}}^{(L)}=\left(-1\right)^{L+m}\sqrt{\left(L+m\right)!\left(L-m\right)!\left(L+m_{1}\right)!\left(L-m_{1}\right)!}\\
\times\sum_{k}\left(-1\right)^{k}\frac{\left(\cos\frac{\theta}{2}\right)^{2k-m-m_{1}}\left(\sin\frac{\theta}{2}\right)^{2L+m+m_{1}-2k}}{k!\left(L+m-k\right)!\left(L+m_{1}-k\right)!(k-m-m_{1})!}
\end{split}
\end{equation}
In the limit of small angle $\theta$ this expression can be simplified up to the linear terms in $\theta$
\begin{equation}\label{eq:d-matrix-linear}
\begin{split}
   d_{mm_{1}}^{(L)}\approx \frac{\left(-1\right)^m\left(-1\right)^{(m+m_1)/2}\sqrt{\left(L+m\right)!\left(L-m\right)!\left(L+m_{1}\right)!\left(L-m_{1}\right)!}}{\left(L+(m+m_{1})/2\right)!\left(m/2-m_{1}/2\right)!\left(m_{1}/2-m/2\right)!\left(L-(m+m_{1})/2\right)!}+\\\frac{\left(-1\right)^m\left(-1\right)^{(m+m_1-1)/2}\sqrt{\left(L+m\right)!\left(L-m\right)!\left(L+m_{1}\right)!\left(L-m_{1}\right)!}}{\left(L+(m+m_{1}-1)/2\right)!\left(m/2-m_{1}/2+1/2\right)!\left(m_{1}/2-m/2+1/2\right)!\left(L-(m+m_{1}+1)/2\right)!}\frac{\theta}{2}
   \end{split}
\end{equation}
The first line of the last expression is different from zero only for $m=m_{1}$, while the second one is non-zero for $m-m_{1}=\pm 1$. It is clear from this expression that the linear in $\theta$ part of the $d_{mm_1}^{(L)}$ is antisymmetric as expected:  $d_{mm_1}^{(L)}=-d_{m_1m}^{(L)}$. 
Thus, we can write
\begin{equation}
    d_{mm_{1}}^{(L)}=\delta_{mm_{1}}+\left(\delta_{m,m_{1}-1}\sqrt{\frac{\left(L-m\right)!\left(L+m+1\right)!}{\left(L+m\right)!\left(L-m-1\right)!}}-\delta_{m,m_{1}+1}\sqrt{\frac{\left(L+m\right)!\left(L-m+1\right)!}{\left(L-m\right)!\left(L+m-1\right)!}}\right)\frac{\theta}{2}
\end{equation}
We will assume now that the $Z$-axis of the stationary coordinate system lies in the equatorial plane of the resonator (defined by the fundamental WGM), while its $Y$-axis is perpendicular to the plane \cite{Deych2009, Rubin2010}. We further assume that the equilibrium position of the oscillator is at the point with coordinates $(0,0,z_{0})$ and that the oscillator, characterized by mass $M$ and frequency $\Omega$, is allowed to move only in $Y-Z$ plane. Assuming small deviations of the particle from its equilibrium position, we can relate particle's spherical coordinates $r,\:\theta$ to its Cartesian displacements $q_y$ and $q_z$ in the $Y$ and $Z$ directions: 
\begin{equation}\label{eq:distance-linear}
    \begin{split}
        r	\approx	r_{0}+q_z\\
        \theta	\approx	-\frac{q_y}{r_{0}}
    \end{split}
\end{equation}
where $r_{0}=z_{0}$. In this approximation the effects of the oscillations are separated: the displacement in the $Z$- direction affects only the frequency of the WGM, while the displacement in the orthogonal direction is responsible for coupling between different WGMs.

Substituting Eq \ref{eq:distance-linear} into Eq.\ref{eq:d-matrix-linear}, we can write 
\begin{equation}
    d_{mm_{1}}^{(L)}=\delta_{mm_{1}}-\left(\delta_{m,m_{1}-1}\sqrt{\frac{\left(L-m\right)!\left(L+m+1\right)!}{\left(L+m\right)!\left(L-m-1\right)!}}-\delta_{m,m_{1}+1}\sqrt{\frac{\left(L+m\right)!\left(L-m+1\right)!}{\left(L-m\right)!\left(L+m-1\right)!}}\right)\frac{q_y}{2r_{0}}
\end{equation}
The optical Hamiltonian in this approximation along with the optomechanical interaction can now be written as 
\begin{equation}
    \begin{aligned}
\hat{H}_{opt}&=\sum_{m}\hbar\omega_{L,m}\left(r_{0}\right)\hat{a}_{L,m}^{\dagger}\hat{a}_{L,m}+\sum_{m}\hbar\frac{d\omega_{L,m}}{dr}\hat{a}_{L,m}^{\dagger}\hat{a}_{L,m}\hat{q_z}+\\
&i\frac{\hat{q}_y}{2r_{0}}\left[\sum_{m}\hbar\omega_{L,m}\left(r_{0}\right)\sqrt{\frac{\left(L-m\right)!\left(L+m+1\right)!}{\left(L+m\right)!\left(L-m-1\right)!}}\left(\hat{a}_{L,m}^{\dagger}\hat{a}_{L,m+1}-\hat{a}_{L,m+1}^{\dagger}\hat{a}_{L,m}\right)+\right.\\
&\left.\sum_{m}\hbar\omega_{L,m}\left(r\right)\sqrt{\frac{\left(L-m+1\right)!\left(L+m\right)!}{\left(L+m-1\right)!\left(L-m\right)!}}\left(\hat{a}_{L,m}^{\dagger}\hat{a}_{L,m-1}-\hat{a}_{L,m-1}^{\dagger}\hat{a}_{L,m}\right)\right]
\end{aligned}
\end{equation}
Incorporating
\begin{eqnarray}
\hat{q}_y	=	\sqrt{\frac{\hbar}{2M\Omega}}\left(\hat{b}_{y}^{\dagger}+\hat{b}_{y}\right)\\
\hat{q}_z	=	\sqrt{\frac{\hbar}{2M\Omega}}\left(\hat{b}_{z}^{\dagger}+\hat{b}_{z}\right)
\end{eqnarray}    
where we introduced standard  creation-annihilation operators for the oscillator:  
\[[b_j,b_k^\dagger]=[a_j,a_k^\dagger]=\delta_{j,k}\]
we arrive at the Hamiltonian along with the mechanical oscillator with frequency $\Omega$
\begin{equation}\label{eq:Hamiltonian-many-modes}
\begin{aligned}
    \hat{H} &= \sum_{m} \hbar \omega_{L,m}(r_{0}) \hat{a}_{L,m}^{\dagger} \hat{a}_{L,m} 
    + \hbar \Omega \left(\hat{b}_{y}^{\dagger} \hat{b}_{y} + \hat{b}_{z}^{\dagger} \hat{b}_{z} \right)
     + \hbar \sum_{m} g_{m}^{(z)}\hat{a}_{L,m}^{\dagger} \hat{a}_{L,m} (\hat{b}_{z}^{\dagger} + \hat{b}_{z}) \\
    &\quad + \frac{i}{2} (\hat{b}_{y}^{\dagger} + \hat{b}_{y}) \Bigg[ 
    \sum_{m} \hbar g_{m}^{(y)} 
    \sqrt{ \frac{(L-m)! (L+m+1)!}{(L+m)! (L-m-1)!} } 
    \left( \hat{a}_{L,m}^{\dagger} \hat{a}_{L,m+1}  - \hat{a}_{L,m+1}^{\dagger} \hat{a}_{L,m} \right) \\
    &\quad + \sum_{m} \hbar g_{m}^{(y)} 
    \sqrt{ \frac{(L-m+1)! (L+m)!}{(L+m-1)! (L-m)!} } 
    \left( \hat{a}_{L,m-1}^{\dagger} \hat{a}_{L,m} - \hat{a}_{L,m}^{\dagger} \hat{a}_{L,m-1} \right)
    \Bigg]
\end{aligned}
\end{equation}
where we also included the mechanical part of the Hamiltonian and introduced optomechanical coupling coefficients $g_m^{(z)}$ and $g_m^{(y)}$ defined as 
\begin{eqnarray}
g_{m}^{(z)}	=	\sqrt{\frac{\hbar}{2M\Omega}}\frac{d\omega_{L,m}}{dr}\\
g_{m}^{(y)}	=	\sqrt{\frac{\hbar}{2M\Omega}}\frac{\omega_{L,m}}{r_{0}}
\end{eqnarray}
Limiting summation over m only to values $-1,0,1$, we can rewrite Hamiltonian of Eq.\ref{eq:Hamiltonian-many-modes} as 

\begin{equation}\label{eq:Hamiltonian-3-mode-supp}
    \begin{split}
        \hat{H}=&\hbar  \omega_{1}\left(\hat{a}_{1}^{\dagger}\hat{a}_{1}+\hat{a}_{-1}^{\dagger}\hat{a}_{-1}\right)+\hbar \omega_{0}\hat{a}_{0}^{\dagger}\hat{a}_{0}+\hbar\Omega\left(\hat{b}_{y}^{\dagger}\hat{b}_{y}+\hat{b}_{z}^{\dagger}\hat{b}_{z}\right)\\ &
+\hbar g_{0}^z\left(\hat{b}_{z}^{\dagger}+\hat{b}_{z}\right)\left(\hat{a}_{0}^{\dagger}\hat{a}_{0}\right)+\hbar g_{1}^z\left(\hat{b}_{z}^{\dagger}+\hat{b}_{z}\right)\left(\hat{a}_{1}^{\dagger}\hat{a}_{1}+\hat{a}_{-1}^{\dagger}\hat{a}_{-1}\right)\\
    &+\frac{i}{2}\hbar g_y \left(\hat{b}_{y}^{\dagger}+\hat{b}_{y}\right)\left[\left(\hat{a}_{-1}^{\dagger}+\hat{a}_{1}^{\dagger}\right)\hat{a}_{0}-\hat{a}_{0}^{\dagger}\left(\hat{a}_{-1}+\hat{a}_{1}\right)\right].
    \end{split}
\end{equation}
where we dropped index $L$ to simplify notation and introduced simplified notation for the optomechanical coupling coefficients
\begin{equation}\label{eq:coupling-zy}
    \begin{split}
        &g_{1,0}^{z}	=	x_{zpf}\frac{d\omega_{L,(1,0)}}{dr}\\
        &g_y	=-L	x_{zpf}\frac{\Delta}{r_{0}}\quad; \Delta=\omega_0-\omega_1 
    \end{split}
\end{equation}
Since the zero-quantum-fluctuation factor $x_{xpf}$:
$$
x_{zpf}=\sqrt{\frac{\hbar}{2M\Omega}}
$$
has been included in the definition of the optomechanical coupling coefficients $g^{z,y}$ it is convenient to redefine  coordinate  operators as dimensionless quantities
\begin{equation}
\hat{q}_{y,z}	=	\frac{\hat{b}_{y,z}^{\dagger}+\hat{b}_{y,z}}{\sqrt{2}}.
\end{equation}
These are the operators, which we use in the main paper, where to avoid introducing extra symbols we keep  the same notation for dimensional and dimensionless quantities.  
\section{Modulation Frequency}

To describe the actual dynamics of the system, we move from a quantum to a classical description by replacing all operators with their expectation values and omitting the hat notation. We then linearize the resulting equations of motion around the corresponding fixed point. In this way, the dynamics of the coupled mechanical mode $q_y$ and the optical degrees of freedom can be written in matrix form. 
\begin{equation}\label{eq:linearized}
    \begin{aligned}
        \frac{d}{dt}
\begin{pmatrix}
q_y  \\
p_y  \\
J_x  \\
J_y
\end{pmatrix}=
        \begin{bmatrix}
        0 & 1 & 0 & 0\\
        -1 & 0 & 0 & 2 g_y\\
        -2g_yJ_z^{eq} & 0 & 0 & -\Delta\\
        0 & 0 & \Delta & 0
        \end{bmatrix}
        \begin{pmatrix}
q_y  \\
p_y  \\
J_x  \\
J_y
\end{pmatrix}
    \end{aligned}
\end{equation}
where the time all quantities of the dimension of frequency are normalized by $1/\Omega$ or $\Omega$ respectively. Within this linearized approximation, the mechanical displacement $q_z$ and the longitudinal optical component $J_z$ decouple from the transverse dynamics. The eigenfrequencies of this linear system are $\pm i\lambda$ and $\pm i\nu$, defined by 
\begin{equation}\label{eq:eigenfreq-supp}
        \lambda	=\sqrt{\frac{(\Delta^{2}+1)+\sqrt{D}}{2}}\:;\:
        \nu	=	\sqrt{\frac{(\Delta^{2}+1)-\sqrt{D}}{2}},
\end{equation}
where 
$$
D=(\Delta^{2}-1)^2-8g_y^{2}N_b\Delta,
$$
and we took in to account that  $J_z^{eq}=N_b/2$.  Transforming the system of equation to the basis of the eigenmodes of the linear system and introducing normal mode coordinates 
\begin{equation}
\begin{pmatrix}
\psi_1\\
\psi_2\\
\psi_3\\
\psi_4 
\end{pmatrix}= U^{-1}\begin{pmatrix}
    q_y\\
    p_y\\
    J_x\\
    J_y
\end{pmatrix},
 \end{equation}
where $U$ is the transformation matrix formed by the (non-normalized) eigenvectors of Eq.\ref{eq:linearized}:
\begin{equation}
    \begin{pmatrix}
    -\frac{\nu^2}{\Delta g_y} & -\frac{\nu^2}{\Delta g_y} & -\frac{\lambda^2}{\Delta g_y}& -\frac{\lambda^2}{\Delta g_y} \\
    i\frac{\nu^2\lambda}{\Delta g_y} & -i\frac{\nu^2\lambda}{\Delta g_y} & i\frac{\nu\lambda^2}{\Delta g_y} & -i\frac{\nu\lambda^2}{\Delta g_y} \\
    -i\frac{\lambda}{\Delta} & i\frac{\lambda}{\Delta} & -i\frac{\nu}{\Delta} & i\frac{\nu}{\Delta}\\
    1 & 1 & 1 & 1
    \end{pmatrix}
\end{equation}
we obtain
\begin{equation}
    \begin{aligned}
        \frac{\textbf{d}}{\textbf{d}\tau}
\begin{pmatrix}
\psi_1  \\
\psi_2  \\
\psi_3  \\
\psi_4
\end{pmatrix}
=
\begin{pmatrix}
-i\lambda & 0        & 0     & 0 \\
0         & i\lambda &  0    & 0 \\
0         & 0        & -i\nu & 0 \\
0         & 0        & 0     & i\nu 
\end{pmatrix}
\begin{pmatrix}
\psi_1  \\
\psi_2  \\
\psi_3  \\
\psi_4
\end{pmatrix}
+
U^{-1}MU
\begin{pmatrix}
\psi_1  \\
\psi_2  \\
\psi_3  \\
\psi_4
\end{pmatrix},
\end{aligned}
\end{equation}
where
\begin{equation}
M=
\begin{bmatrix}
0 & 0 & 0 & 0\\
0 & 0 & 0 & 0\\
-2g_y\Delta J_z & 0 & \sqrt{2}(g_1^z-g_0^z) q_z &0\\
0 & 0 & 0 & -\sqrt{2}(g_1^z-g_0^z) q_z,
\end{bmatrix}
\end{equation}
where $\Delta J_z=J_z-J_z^{eq}$. It is easy to see that $\psi_1=\psi_2^*$ and $\psi_3=\psi_4^*$.

Assuming that  $\Delta \gg 1$ and that the system is far from the bifurcation point   $\Delta^3 \gg 8g_y^2N_b$ we can approximate the eigenvalues as 
\begin{equation}
\lambda \approx \Delta \:;\: \nu \approx 1
\end{equation}

It can be shown that in this approximation the contribution of the nonlinearity to the dynamics of $\psi_{3,4}$ can be completely neglected while equations for  $\psi_{1,2}$ can be significantly simplified
\begin{equation}
   \begin{aligned}
    \dot{\psi}_1&=i\lambda\psi_1+i\Delta J_z{\lambda}(\psi_3+\psi_3^*) \\
    \dot{\psi}_3&=i\nu\psi_3
    \end{aligned} 
\end{equation}
where we, taking into account our numerical results, also neglected terms with $q_z$. 

These equations are convenient to rewrite in terms of the real and imaginary parts of the modal amplitudes:
\begin{equation}
    \begin{aligned}
        \frac{d\eta_1^+}{dt}=&-\lambda\eta_1^-\\
        \frac{d\eta_1^-}{dt}=&\lambda\eta_1^++2{\lambda} \Delta J_z\eta_2^+\\
        \frac{d\eta_2^+}{dt}=&-\nu\eta_2^-\\
        \frac{d\eta_2^-}{dt}=&\nu\eta_2^+\\
        \frac{d \Delta J_z}{dt}=&2{\lambda} \eta_1^-\eta_2^+
    \end{aligned}
\end{equation}
defined as 
\begin{equation}
\begin{aligned}
\eta_1^+&=&(\psi_1+\psi_1^*)/2 \\
\eta_1^-&=&(\psi_1-\psi_1^*)/2i \\
\eta_2^+&=&(\psi_3+\psi_3^*)/2 \\
\eta_2^-&=&(\psi_3-\psi_3^*)/2i
\end{aligned}
\end{equation}
Equations for $\eta_2^\pm$ can be immediately solved, while variables $\eta_1^\pm$ can be presented as a product of  the fast oscillatory terms and the slow amplitude modulation: 
\begin{equation}
    \begin{aligned}
        &\eta_1^{+}=\frac{g_y q_0}{2\lambda}(r^+(t)\cos{\lambda t }+ r^-(t)\sin{\lambda t})\\
        &\eta_1^{-}=\frac{g_y q_0}{2\lambda}(r^+(t)\sin{\lambda t }- r^-(t)\cos{\lambda t})
    \end{aligned}
\end{equation}
where $r^{\pm}(t)$ represent slowly varying amplitude functions satisfying initial conditions  $r^{+}(0)=1, r^{-}(0)=0$ and $q_0$ represents the initial perturbation in $y$ direction.

Substituting this ansatz as well solutions for $\eta_2^\pm$ into the normal mode equations and averaging the resulting equations over fast oscillations we obtain the  equations for the slow changing amplitude $r^\pm$:
\begin{equation}
   \ddot{r}^\pm+\frac{\left(g_yq_0\right)^2}{8}r^\pm=0
\end{equation}
describing the modulation of the amplitude with frequency
$$
\omega_{mod}=\frac{g_yq_0}{2\sqrt{2}}
$$
This modulation frequency $\omega_{\text{mod}}$ determines the time scale for energy exchange between the coupled mechanical and optical degrees of freedom.

\section{Adiabatic projection and rotation matrix}

For slowly varying mechanical coordinates \(q_y(t),q_z(t)\) the optical dynamics at fixed instant is a rigid rotation of
\(\mathbf{J}=\{J_x,J_y,J_z\}\) with instantaneous angular velocity
\[
\mathbf{\Lambda}(t)=\big(0,-2\,g_y q_y(t),\,\,-(\Delta-\sqrt{2}(g_1^z-g_0^z) q_z(t))\big),
\qquad
\Lambda(t)=\|\mathbf{\Lambda}(t)\|.
\]
The unit rotation axis and its components are
\begin{equation}\label{eq:n_supp}
\mathbf{n}(t)=\frac{\mathbf{\Lambda}_q(t)}{\Lambda(t)}=(0,n_y,n_z),\quad
n_y=-\frac{2\,g_y q_y}{\Lambda},\quad
n_z=-\frac{\Delta-\sqrt{2}(g_1^z-g_0^z) q_z}{\Lambda}.
\end{equation}
The adiabatic invariant is the axis projection
\begin{equation}\label{eq:u_supp}
u(t)\equiv \mathbf{J}(t)\cdot\mathbf{n}(t)=J_y\,n_y+J_z\,n_z.
\end{equation}

For frozen mechanical variables (constant \(\mathbf{n}\) over the short interval) the optical evolution is
\(\mathbf{J}(t)=\hat R(\theta,\mathbf{n})\,\mathbf{J}_0\) 
where rotation matrix $\hat{R}$ can be expressed in terms components of the unit vector $\mathbf{n}$ as
\begin{equation}\label{eq:R_explicit_short}
\hat R(\theta)=
\begin{pmatrix}
\cos\theta + (1-\cos\theta)n_y^2 & -n_z\sin\theta & (1-\cos\theta)n_y n_z\\[6pt]
n_z\sin\theta & \cos\theta & -n_y\sin\theta\\[6pt]
(1-\cos\theta)n_y n_z & n_y\sin\theta & \cos\theta + (1-\cos\theta)n_z^2
\end{pmatrix}.
\end{equation}
where $\theta$ is the fast variable that can be approximated as $\theta\approx\int dt\Lambda_q (t)$.
\section{Dissipation}
As mentioned in the main text, a detailed analysis of dissipation effects, particularly in conjunction with pulse excitation, lies beyond the scope of the present paper. Nevertheless, in order to demonstrate the robustness of the broad optical sideband spectrum, we performed numerical simulations including both optical and mechanical dissipation. Optical dissipation was introduced in the standard way into the equations for the optical amplitudes $a_0$ and $a_{\pm1}$. These equations were then transformed into equations for the Schwinger operators under the simplifying assumption that the optical decay rates of all modes are identical. As a result, the decay of all components of the pseudospin operator $\mathbf{J}$ is governed by the same parameter $\Gamma_{\mathrm{opt}}$. Mechanical dissipation was introduced in the standard way through the terms $\Gamma_M \dot{q}_{y,z}$.
\begin{equation}\label{eq:EOM}
     \begin{aligned}
    \dot{q}_z &= p_z,  \\
    \dot{p}_z &= -q_z
               - \sqrt{2}\,g_1^z\,\frac{1}{2}\!\left(N-2J_z\right)
               - \Gamma_M\,\dot{q}_z, \\
    \dot{q}_y &= p_y, \\
    \dot{p}_y &= -q_y
               - 2\,g_yJ_y
               - \Gamma_M\,\dot{q}_y, \\
    \dot{J}_x &= -\!\left(\Delta - \sqrt{2}\,g_1^z\,q_z\right)J_y
               + 2\,g_y\,q_y\,J_z - \Gamma_{opt}\,J_x, \\
    \dot{J}_y &= \phantom{-}\!\left(\Delta - \sqrt{2}\,g_1^z\,q_z\right)J_x
               - \Gamma_{opt}\,J_y, \\
    \dot{J}_z &= -2\,g_y\,q_y\,J_x - \Gamma_{opt}\,J_z\\
    \dot{N}&=-2\Gamma_{opt}N
     \end{aligned}
 \end{equation}
The last equation in the set takes into account that the total photon number $N$, which is conserving in the absence of the dissipation is now a decaying function of time. As a result, the expression for the optical energy, Eq. 49 for the energy of the main text must be modified taking into account that term $\hbar\omega_0N$ is no longer constant and cannot be omitted: 
\begin{equation}\label{eq:en_diss}
    E_{opt}=\frac{1}{2}\left(\omega_0+\omega_1\right)N-\left(\omega_1-\omega_0\right)J_z
\end{equation}
As a result, in the presence of optical losses, the total optical energy contains a large contribution proportional to the total photon number $N(t)$. Under the assumption of identical decay rates for all optical modes, this contribution is independent of the internal nonlinear dynamics and represents a known exponentially decaying background. Since the broad optical sideband spectrum is associated with the residual  signal proportional to $J_z$, it is convenient to analyze the background-subtracted quantity
\begin{equation}\label{eq:energy-min-back}
\tilde{E}_{\mathrm{opt}}(t)=E_{\mathrm{opt}}(t)-\frac{1}{2}\left(\omega_0+\omega_1\right)N(t)=\left(\omega_1-\omega_0\right)J_z.
\end{equation}
Experimentally, the corresponding background can be determined independently from a reference measurement in which the transverse coupling is absent or negligible and subsequently subtracted from the measured signal. Such a procedure removes the trivial decay associated with photon losses while preserving the nonlinear dynamics responsible for the formation of the broad sideband spectrum.

The corresponding results are shown in the plots below. One can see that the broad sideband structure remains robust for optical decay rates up to approximately $0.1\Omega_M$, while substantial spectral degradation occurs only when $\Gamma_{\mathrm{opt}}\sim\Omega_M$. Even in this regime, the lower-frequency sidebands are suppressed first, whereas the higher-frequency components remain visible. This behavior is physically natural since dissipation affects the slower dynamical components of the system more strongly than the faster ones.
\begin{figure}[H]
    \centering
     \begin{subfigure}{\textwidth}
         \centering     \includegraphics[width=\textwidth]{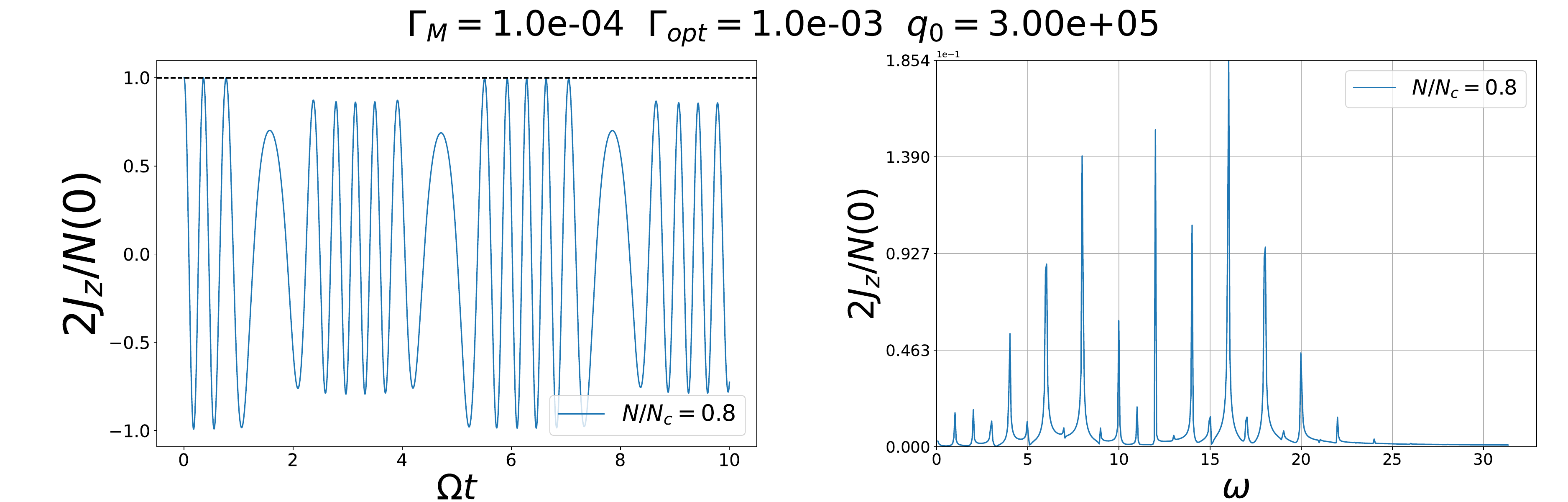}
         \end{subfigure}
         \vfill
     \begin{subfigure}{\textwidth}
         \centering         \includegraphics[width=\textwidth]{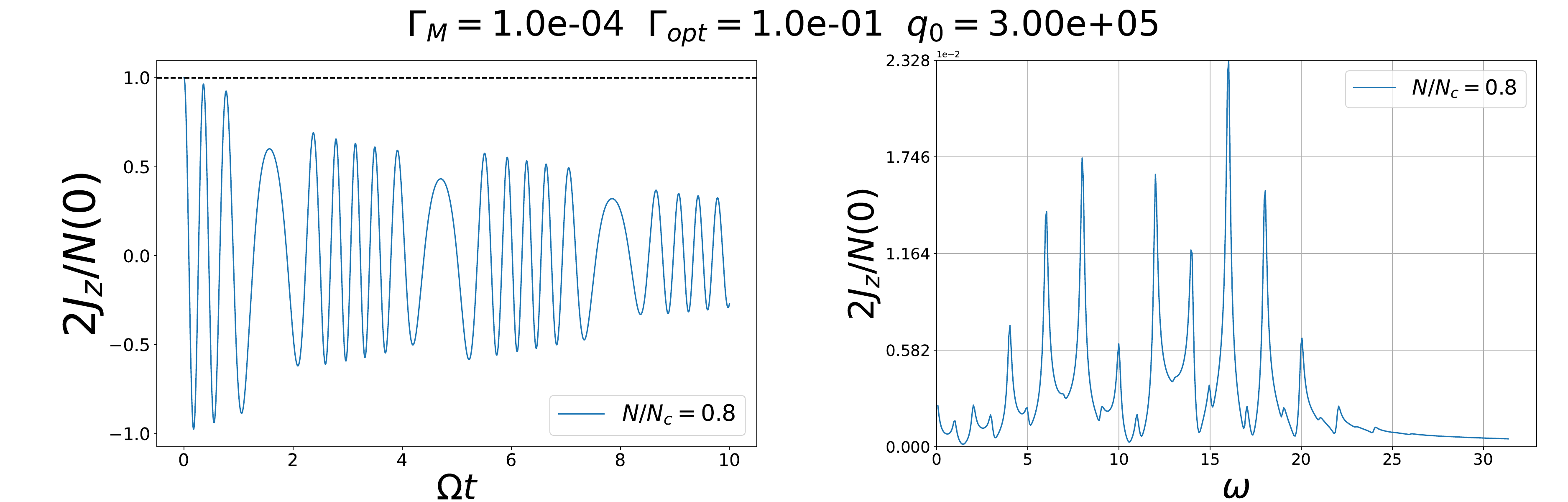}
   \end{subfigure}   
         \vfill
     \begin{subfigure}{\textwidth}
         \centering         \includegraphics[width=\textwidth]{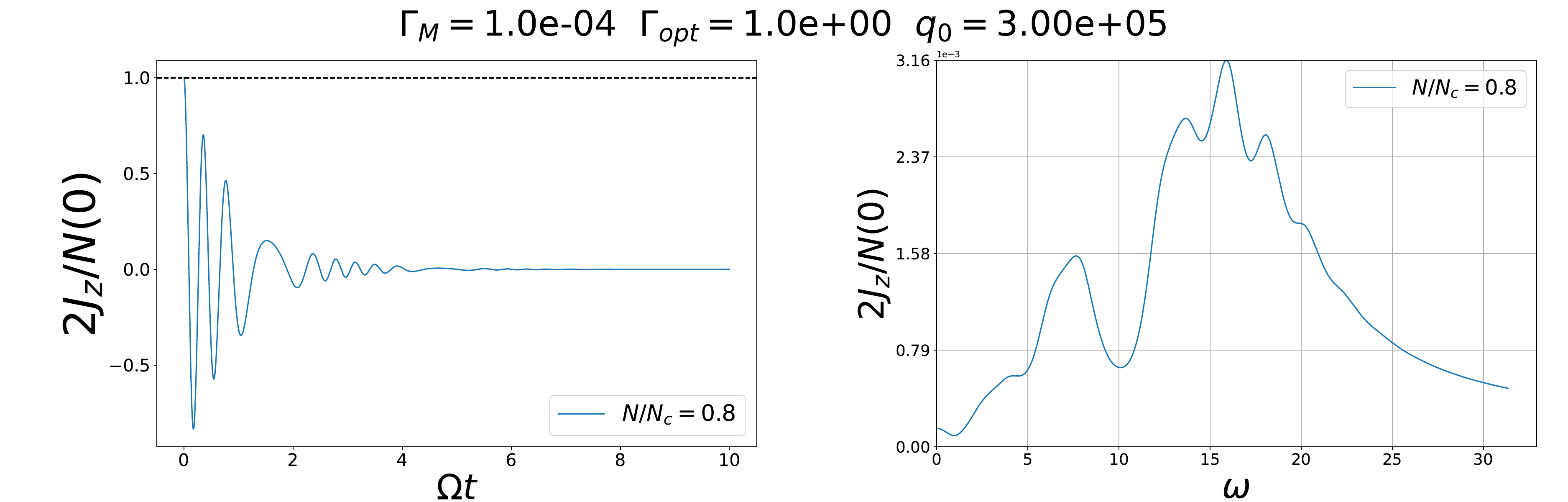}
   \end{subfigure}  
     \caption{Evolution of optical energy with subtracted trivial background, Eq.\ref{eq:energy-min-back} (top)  and the corresponding Fourier spectra (bottom) for $N(0)=0.8N_c$ for strong non-linear case $q_0=3.00\times 10^5$ for various optical decay parameters $\Gamma_{opt}=[10^{-3},10^{-1},1]$ and $\Gamma_M=10^{-4}$. $N(0)$ is the initial value of the photon number in the cavity. } 
\end{figure}

\section{Numerical Integration Techniques}
 
In this section we provide details of our numerical analysis and the respective code. The coupled dynamics of the spin-mechanical system are
described by a set of seven first-order ordinary differential equations
(ODEs) for the dimensionless mechanical quadratures
$(q_z,\,p_z,\,q_y,\,p_y)$ and the collective Bloch-vector components
$(J_x,\,J_y,\,J_z)$, Eq.\ref{eq:EOM}
These equations are integrated numerically
using the explicit, adaptive-step Runge–Kutta~45 (Dormand–Prince) method
\cite{dormand1980rk45} as implemented in \texttt{scipy.integrate.solve\_ivp}
\cite{virtanen2020scipy}.
The integration is performed over the time window $[0,\,t_f]$ and the
solution is evaluated at the predefined time grid \texttt{t\_eval}.
Absolute and relative error tolerances are set to
$\varepsilon_{\mathrm{abs}}=10^{-9}$ and
$\varepsilon_{\mathrm{rel}}=10^{-7}$, respectively, ensuring sub-per-mil
accuracy throughout the trajectory.
The corresponding Python implementation is given in
Listing~\ref{lst:spinode}.
\begin{figure}[H]
\begin{minipage}{\linewidth}
\begin{verbatim}
def spin_ODE(t, x, N):
    q_z, p_z, q_y, p_y, Jx, Jy, Jz, n = x
 
    q_z_dot = p_z
    p_z_dot = -q_z - 2**0.5*g_z*(N/2)*(n - Jz) - Gamma_M*p_z
    q_y_dot = p_y
    p_y_dot = -q_y - 2*g_y*(N/2)*Jy               - Gamma_M*p_y
 
    Jx_dot = -(Delta - 2**0.5*g_z*q_z)*Jy + 2*g_y*q_y*Jz - Gamma_opt*Jx
    Jy_dot =  (Delta - 2**0.5*g_z*q_z)*Jx                - Gamma_opt*Jy
    Jz_dot = -2*g_y*q_y*Jx                               - Gamma_opt*Jz
    n_dot=-2*Gamma_opt*n
    return [q_z_dot, p_z_dot, q_y_dot, p_y_dot, Jx_dot, Jy_dot, Jz_dot,n_dot]
 
xx = solve_ivp(
    spin_ODE,            # callable f(t, x, *args)
    [0, tf],             # integration interval
    x0,                  # initial state vector (length 8)
    args=[N],            # extra positional arguments passed to spin_ODE
    t_eval=t_eval,       # output time grid
    dense_output=False,  # disable continuous solution object
    rtol=1e-7,           # relative error tolerance
    atol=1e-9,           # absolute error tolerance
    method="RK45"        # Dormand-Prince explicit Runge-Kutta
)
\end{verbatim}
\caption{Python implementation of the ODE and its
    numerical integration with \texttt{scipy.integrate.solve\_ivp}.}
\label{lst:spinode}
\end{minipage}
\end{figure}
 
\noindent
The call signature of $\texttt{solve\_ivp}$ and the role of each argument are
as follows.
The first positional argument is the ODE function $\texttt{spin\_ODE}$, which
accepts the current time $t$, the state vector $\mathbf{x}$, and the
additional parameter $N$; it returns $\dot{\mathbf{x}}$ as a plain Python
list.
The second argument specifies the integration interval $[0, t_f]$.
The initial condition $\texttt{x0}$ is a length-8 array containing
$(q_z, p_z, q_y, p_y, J_x, J_y, J_z)|_{t=0}$.
The keyword argument $\texttt{args}$ forwards $N$ to $\texttt{spin\_ODE}$ as
an extra positional parameter.
Setting $\texttt{dense\_output=False}$ suppresses the construction of a
continuous interpolant, reducing memory overhead when only discrete output
samples are needed.
The time points at which the solution is stored are supplied via
$\texttt{t\_eval}$; this does not constrain the internal adaptive step-size
selection.
The $\texttt{RK45}$ method automatically adjusts the step size to keep the
local truncation error below the prescribed tolerances, making it
well-suited for the moderately stiff spin-mechanical dynamics encountered
here.
The solution object $\texttt{xx}$ exposes the trajectory as
$\texttt{xx.y}$ (shape $8\times|\texttt{t\_eval}|$) and the corresponding
times as $\texttt{xx.t}$.